\documentclass[fleqn,usenatbib]{mnras}

\usepackage{newtxtext,newtxmath}

\usepackage[T1]{fontenc}

\DeclareRobustCommand{\VAN}[3]{#2}
\let\VANthebibliography\thebibliography
\def\thebibliography{\DeclareRobustCommand{\VAN}[3]{##3}\VANthebibliography}

\usepackage{graphicx}	
\usepackage{amsmath}	

\usepackage{amssymb}	
\usepackage{multirow}
\usepackage{booktabs}
\usepackage[flushleft]{threeparttable}
\usepackage{caption}
\usepackage{hyperref}
\usepackage{ulem}
\usepackage{fix-cm}
\newcommand{\orcid}[1]{\href{https://orcid.org/#1}{\includegraphics[width=8pt]{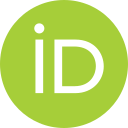}}}

\title[Lunar-based VLBI BH shadow detection]{Beyond Sgr A* and M87*: Sub-Microarcsecond Black Hole Shadow Detection via Lunar-based Extremely Long Baseline Interferometry}

\author[S. Zhao et al.]{
Shan-Shan Zhao,$^{\orcid{0000-0002-9774-3606},1}$\thanks{E-mail: zhaoss@shao.ac.cn}
Ru-Sen Lu$^{\orcid{0000-0002-7692-7967},1,2,3}$,
Lei Liu$^{\orcid{0000-0002-2920-1880},1}$,
Zhiqiang Shen$^{\orcid{0000-0003-3540-8746},1,2}$,
and Yosuke Mizuno $^{\orcid{0000-0002-8131-6730},4,5,6,7}$
\\
$^{1}$Shanghai Astronomical Observatory, Chinese Academy of Sciences, Shanghai, 200030, P. R. China;\\
$^{2}$Key Laboratory of Radio Astronomy and Technology, Chinese Academy of Sciences, A20 Datun Road, Chaoyang District, Beijing, 100101, P. R. China;\\
$^{3}$Max-Planck-Institut für Radioastronomie, Auf dem Hügel 69, D-53121 Bonn, Germany;\\
$^{4}$Tsung-Dao Lee Institute, Shanghai Jiao Tong University, 1 Lisuo Road, Shanghai 201210, P. R. China;\\
$^{5}$School of Physics and Astronomy, Shanghai Jiao Tong University, 800 Dongchuan Road, Shanghai 200240, P. R. China;\\
$^{6}$Key Laboratory for Particle Physics, Astrophysics,  and Cosmology (MOE), Shanghai Key Laboratory for Particle Physics and Cosmology, \\Shanghai Jiao Tong University, 800 Dongchuan Road, Shanghai 200240, P. R. China;\\
$^{7}$Institut f\"ur Theoretische Physik, Goethe-Universit\"at, Max-von-Laue-Strasse 1, D-60438, Frankfurt am Main, Germany
}

\date{Accepted XXX. Received YYY; in original form ZZZ}

\pubyear{2024}

\begin{document}
\label{firstpage}
\pagerange{\pageref{firstpage}--\pageref{lastpage}}
\maketitle

\begin{abstract}
The 1.3 mm ground-based very long baseline interferometry (VLBI) array Event Horizon Telescope (EHT), is limited by Earth's diameter, restricting horizon-scale imaging to only M87* and Sgr~A*. Extending baselines to the Moon would achieve $\sim0.7\ \mu$as angular resolution at 230\,GHz, enabling shadow detection for many more supermassive black holes (SMBHs). The concept is motivated by space VLBI missions and lunar exploration, including the ongoing Lunar Orbit VLBI EXperiment (LOVEX) aboard QueQiao-2 (Chang’E-7) and the planned International Lunar Research Station (ILRS). We assess shadow detectability for 31 SMBHs with predicted large angular sizes, assuming optically thin emission at 230 GHz, exploring different telescope locations and antenna sizes. Assuming a telescope at the lunar antipode, we simulate the Moon–Earth ($u,v$) coverage and show that sources near the Moon's orbital plane yield projected baselines spanning a wide range, enabling sampling of the first visibility null - a key shadow signature. Using a geometric ring model, we identify six shadow-detectable candidates: M104, NGC~5077, and NGC~1052 are detectable with a 5 m lunar-based telescope; PGC~049940 with 10 m; NGC~524 with 20 m; and NGC~5252 with 40 m. If additional space telescopes fill the baseline coverage gaps between Moon and Earth, 14 candidates are detectable for the $n=1$ photon-ring region with a lunar-based telescope up to 40 m. These results provide a clear scientific and technical motivation for lunar-based telescopes in future black hole shadow studies. 

\end{abstract}




\section{Introduction}

Images of the black hole shadows of the supermassive black holes (SMBHs) at the centers of M87 (M87*) and the Milky Way (Sgr A*) have been obtained with an angular resolution of $\sim20~\mu$as using the Event Horizon Telescope (EHT), a global 230\,GHz very long baseline interferometry (VLBI) array \citep{EHTC2019ApJL875.L1,EHTC2022ApJ930L12,EHTC2024AA681.79}. The angular resolution of VLBI scales as the observing wavelength divided by the baseline length and can therefore be improved by observing at higher frequencies or by extending the baselines. The next-generation EHT (ngEHT) plans to operate at 345\,GHz, potentially improving the resolution to $\sim15~\mu$as \citep{Doeleman2019BAAS51.256,Johnson2023Galax11.61}, and 345\,GHz VLBI has been successfully tested \citep{Raymond2024AJ168.130}.
However, atmospheric opacity limits further increasing observing frequencies for ground-based arrays. Reaching the microarcsecond and sub-microarcsecond regimes requires baselines far exceeding Earth’s diameter ($D_\oplus$). For comparison, a geostationary orbit radius of 42,164 km (3.3\,$D_\oplus$) provides a nominal angular resolution of $\sim$6 $\mu$as at 230\,GHz, whereas the mean Earth–Moon distance of 384,400 km (30.1\,$D_\oplus$) corresponds to a baseline resolution of $\sim$0.7 $\mu$as at the same frequency. Consequently, extremely long baseline interferometry incorporating a lunar-based telescope offers a promising pathway toward sub-microarcsecond angular resolution and the direct detection of black hole shadows beyond Sgr A* and M87* (see e.g. \citealt{Falcke2000ApJ528L13} for the black hole shadow concept)

Sub-microarcsecond black hole imaging will profoundly advance tests of Einstein's General Relativity (GR) in strong gravitational fields, enhancing both its precision and universality.
For M87* and Sgr~A*, the key science goal of such high-resolution imaging is to detect fine structures such as photon rings \citep[e.g.,][]{Johnson2020SciA6.1310}, which arise directly from strong gravitational lensing and provide exceptionally stringent tests of GR.
Equally important is extending such tests to a broader and more diverse sample of SMBHs. M87* \citep[$6.5\times10^9\ \rm{M}_{\odot}$;][]{EHTC2019ApJL875.L6} and Sgr A* \citep[$4\times10^6\ \rm{M}_{\odot}$;][]{GRAVITY2022AA657.12} differ in mass by over three orders of magnitude, highlighting the lack of observational coverage for SMBHs with a mass between them. 
The number of resolvable black hole shadows at $\sim 1\ \mu$as resolution is predicted to reach hundreds \citep{Pesce2021ApJ923.260}, and black holes with masses $\gtrsim 10^9\ \rm{M}_\odot$ could in principle be resolved at any redshift with Earth–L2 or larger baselines \citep{Pesce2019BAAS51.176}.

Decades of technical development in space VLBI have brought the field to a level of maturity that enables its application to black hole imaging in the near future. Early tests began with the Tracking and Data Relay Satellite System (TDRSS) in 1986  \citep{Levy1986Sci234.187}, followed by the first dedicated space VLBI mission, the VLBI Space Observatory Programme (VSOP), in 1997 \citep{Hirabayashi1998Sci281.1825}. RadioAstron, launched in 2011, extended baselines up to $\sim 28\ D_\oplus$ \citep{Kardashev2013ARep57.153}, achieving imaging resolutions of 12 $\mu$as for OJ 287 \citep{Gomez2022ApJ924.122} and 150 $\mu$as for M87* \citep{Kim2023ApJ952.34} at 22\,GHz, although fringes were not detected for Sgr~A* \citep{Johnson2021ApJ922.28}. 

Future missions target millimeter/submillimeter black hole imaging. Black Hole Explorer (BHEX) \footnote{\url{https://www.blackholeexplorer.org/}} proposes a 3.5-meter telescope in medium-Earth orbit (MEO) forming a space-Earth VLBI array with EHT/ngEHT to capture photon rings \citep{Johnson2024SPIE13092E..2DJ,Marrone2024SPIE13092E..2GM}. The Millimetron space observatory (next-generation RadioAstron), proposes a millimeter telescope at Sun-Earth L2 to form space-ground VLBI arrays  \citep{Likhachev2022MNRAS511.668,Andrianov2021MNRAS500.4866}. Space-space VLBI is also a promising concept. TeraHertz Exploration and Zooming-in for Astrophysics (THEZA), including Event Horizon Imager (EHI)/Space-based High-resolution Array for Radio astronomy and Physics (SHARP), envisions three MEO satellites and uses their differing orbital periods to form a extremely dense ($u,v$) coverage\citep{Roelofs2019AA625.124,Kudriashov2021ChJSS41.211,Gurvits2021ExA51.559}. CAPELLA plans four low-Earth orbit (LEO) satellites operating as a 690\,GHz space-space interferometer \citep{Trippe2023arXiv2304.06482}. 

Interferometry at Moon-Earth scale represents a unique frontier of space VLBI. Its development has been strongly motivated by lunar exploration needs, particularly precise deep-space orbit determination.
As a first step toward this capability, the Lunar Orbit VLBI EXperiment (LOVEX) project in Chang'E 7 mission has successfully demonstrated Moon-Earth VLBI fringe detections trough the observation of AO~0235+164 at 8\,GHz \citep{Hong2025SCPMA..6919511H}. It operates a 4.2 m radio telescope on the Moon-orbiting satellite QueQiao-2, launched in 2024 and designed to support eight years of observations \citep{Wang2024NSRev11D.329,Zhang2024CSST44.5}. With rapid advances in lunar exploration, establishing a lunar-based astronomical observatory is becoming increasingly feasible. An essential target of China lunar exploration following the Chang'E 8 mission is the construction of the International Lunar Research Station (ILRS), which aims to enable long-term scientific research on the Moon \citep{Li2019Sci365.238}. A possible site for ILRS is in the south pole region \citep{Hu2023PSS227.105623}, and plans include establishing astronomical capabilities during the ILRS-5 mission\footnote{\url{https://www.cnsa.gov.cn/english/n6465652/n6465653/c6812150/content.html}}. 

Motivated by the future space VLBI missions and lunar exploration, we consider a lunar-based telescope at various locations and with different antenna sizes operating jointly with the EHT at 230\,GHz and perform observational simulations. We quantitatively assess the capability of such an array to detect black hole shadows for 31 SMBH candidates with expected large shadow sizes, including M87* and Sgr A*. Several of these candidates have also been proposed for Earth-orbiting space VLBI \citep{Ramakrishnan2023Galax11.15, Zhang2025ApJ985.41, BenZineb2024arXiv2412.01904, Fish2020AdSpR65.821} We focus on 230\,GHz as a representative and most mature observing frequency for this study, while the extension to higher frequencies (e.g., 345\,GHz) is left for future work.

The paper is organized as follows. Section \ref{sec:candidates} introduces the SMBH candidates. Section \ref{sec:lunar_telescope} describes the assumptions for the lunar-based telescope. Section \ref{sec:VLBIsimulation} presents the methods and results of Moon-Earth baseline coverage simulations. Section \ref{sec:detectability} analyzes the black hole shadow detectability. Finally, Section \ref{sec:conclusion} presents the conclusions and discussions. 

The cosmology used in this paper is $H_0 = 71$ km s$^{-1}$ Mpc$^{-1}$, $\Omega_{M}$=0.27, $\Omega_{\Lambda}$=0.73. 

\section{SMBH shadow candidates}
\label{sec:candidates}

\subsection{Basic information}

The selection of candidates for horizon-scale imaging at 230\,GHz is primarily based on the predicted black hole shadow size. This prediction, under the standard assumption of optically thin accretion flows, equates the observed emission ring diameter with the black hole shadow diameter. It generally uses the innermost photon orbit in Schwarzschild spacetime to estimate the shadow diameter, which is \citep[e.g. ][]{Falcke2013CQGra30.24.244003}
\begin{equation}
    \theta=2\sqrt{27}\frac{GM_{\rm BH}}{D_{\rm BH}c^2},\label{eq:ring_size}
\end{equation}
where $G$ is the gravitational constant, $M_{\rm BH}$ and $D_{\rm BH}$ are the mass and distance of the black hole. As a result, black holes with large mass and close distance will have large shadows. 

We can roughly estimate the beam size of a lunar-based telescope paired with the EHT by
\begin{equation}
    \theta_{\rm beam}\sim \dfrac{\lambda}{B_{\rm max}}.\label{eq:theta_beam}
\end{equation}
$\lambda$ is the observing wavelength, which is 1.3 mm for 230\,GHz. $B_{\rm max}$ is the maximum baseline, which can be equal to the distance between Moon and Earth in our case, i.e., 384,400 km. Therefore, $\theta_{\rm beam}\sim0.7\ \mu$as. 

31 selected candidates with predicted shadow size larger than Moon-Earth VLBI beam size ($\theta\gtrsim\theta_{\rm beam}$) are listed in Table~\ref{tab:BHimages_candidates}. 19 of them are suggested to have a ring diameter larger than 1 $\mu$as in Event Horizons and Environs \citep[ETHER, a comprehensive database of candidates for the EHT and ngEHT; ][]{Ramakrishnan2023Galax11.15,Nair2024evn..conf...75N}. Then \cite{Zhang2025ApJ985.41} conducted a detailed analysis of 12 ETHER candidates that are expected to be resolvable by VLBI in space. 14 candidates are analyzed by \cite{BenZineb2024arXiv2412.01904} to determine the optimal configuration of space baselines in high Earth orbits when imaging black holes via space VLBI. 14 candidates are large apparent black holes whose flux density has been successfully detected at 230\,GHz by \cite{Lo2023ApJ950.10}. The above candidate samples contain overlapping sources, which after consolidation yield a total of 31 candidates. We assume all sources are optically thin at 230 GHz, a requirement for visible shadows. While M87* and Sgr~A* do transition from optically thick to optically thin around this frequency \citep{Chael2023ApJ...945...40C}, but for other sources the optical state is less certain, likely depending on accretion and mass \citep{Bandyopadhyay2025AA...698A..24B,Desire2025ApJ...980..262D}; a detailed investigation is left to future work.

Table~\ref{tab:BHimages_candidates} summarizes the basic information of the 31 SMBH candidates, including source name, the location on the celestial sphere, measured 230\,GHz flux density, measured black hole mass and distance, the ring diameter predicted by Eq.~\ref{eq:ring_size}, and the estimated ring width. The ring width is assumed to be 40\% of the ring diameter, consistent with the approximate fraction inferred from EHT observations of Sgr~A* and M87* \citep{EHTC2022ApJ930L12,EHTC2019ApJL875.L1}. The sources are ordered by right ascension, and the corresponding references are listed below the table.

\begin{table*}
\begin{threeparttable}
\caption{SMBH candidates analyzed in this work}
\setlength\tabcolsep{5pt}       
\begin{tabular}{@{}ccccccccccc@{}}
\toprule
Source & Other  & RA            & Dec           & $F_{\rm 230\,GHz}$ & $M_{\rm BH}$   & $D_{\rm BH}$  & $\theta$  & $w$  & Refs. & Ref.Sel.\\
name& name&  & & (Jy) & ($10^8\,\rm{M}_{\odot}$)    & (Mpc) & ($\mu$as)  &  ($\mu$as) & &\\\midrule

Sgr A*               && 17:45:40.0409 & -29:00:28.118 & 2.4   & $0.04^{+0.011}_{-0.006}$ & $8277\pm9\times10^{-6}$   & $51.8\pm2.3$      & $\sim$16-26 & 1, 2, 3 & \\

M87*         &NGC 4486        & 12:30:49.4234  & +12:23:28.044  & 1.2  & $65\pm7$  & $16.8\pm0.8$ & $42\pm3$ & $<20$  & 4, 5, 6&\\\hline

M31         &NGC 224        & 00:42:44.3517  & +41:16:08.673  & 0.0226    & $1.4$  & $0.78$  & 18.4  &  7.4 & 7, 8, 9& L\\

NGC 315              && 00:57:48.8833 & +30:21:08.812 & 0.182  & 20.8      & 70  & 3.0      & 1.2 & 10, 11  & N, Z, B\\

NGC 524              && 01:24:47.7452 & +09:32:20.040 & 0.0182   & 8.32        & 33.6   & 2.5      & 1.0   & 7, 12& L\\

NGC 1052             && 02:41:04.7985 & -08:15:20.752 & 0.35    & 1.54      & 17.6  & 0.9      & 0.4  & 13, 14, 15& N, L\\

NGC 1218       & 3C 78      & 03:08:26.2283    & +04:06:39.300    & 0.11   & 40     & 125  & 3.3      & 1.3  & 16, 17, 18& N, Z, B\\

NGC 1275    &3C 84   & 03:19:48.1601    & +41:30:42.103    & 7.15    & 11     & 62.5  & 1.8      & 0.7 & 19, 20& N \\

NGC 1277    &  & 03:19:51.4902    & +41:34:24.940    & 0.0151    & 49         & 71  & 7.1      & 2.8  & 7, 21& L\\

NGC 1399    &  & 03:38:29.0200    & -35:27:00.700    & 0.0385    & 12        & 19.9  & 6.2      & 2.5  & 7, 22, 23& B, L\\

NGC 1600    &  & 04:31:39.8676    & -05:05:10.587    & 0.0142    & 170         & 64  & 27.2      & 10.9  & 7, 24& L\\

NGC 2663     &        & 08:45:08.1440    & -33:47:41.064    & 0.084  & 6.6     & 27.5  & 2.5      & 1.0  & 25, 26 & N, Z, B\\

M81         &NGC 3031         & 09:55:33.1730    & +69:03:55.060    & 0.1    & 0.7       & 3.96  & 1.8      & 0.7  & 27, 28, 29& N\\

NGC 3894      &       & 11:48:50.3581    & +59:24:56.382    & 0.0576  & 20       & 50.1  & 4.1      & 1.6  & 30, 31& N, Z, B\\

NGC 3998      &       & 11:57:56.1333    & +55:27:12.922    & 0.13   & 8.1      & 14.7 & 5.7      & 2.3  & 25, 32& N, Z, B\\

NGC 4261      &       & 12:19:23.2160    & +05:49:29.700    & 0.32   & 16.7      & 31.1   & 5.5      & 2.2  & 11, 16& N, Z, B\\

M84     &NGC 4374         & 12:25:03.7432    & +12:53:13.138    & 0.116  & 15      & 17   & 9.1      & 3.6 & 10, 12 & N, Z, B\\

NGC 4256 & & 12:34:03.0285 & +07:41:56.904 & 0.0159 & 4.5 & 16.4 & 2.8 & 1.1 & 7, 33 & L\\

M89 & NGC 4552         & 12:35:39.8070    & +12:33:22.831    & 0.0122  & 4.27      & 15.3  & 2.9      & 1.2  & 34& N, Z, B\\

M104& NGC 4594        & 12:39:59.4314    & -11:37:23.118    & 0.198   & 9          & 9.55   & 9.7     & 3.9  & 10, 35, 36& N, Z, B\\

NGC 4751 & & 12:52:50.7623 &	-42:39:35.511&	0.0186 &33.4 & 	26.3 &	13.0 &	5.2 & 7, 37& L\\ 

NGC 5077        &     & 13:19:31.6700    & -12:39:25.076    & 0.120   & 8.0       & 44.9 & 1.8      & 0.7  & 38, 12 & N, Z\\

Cen A* & NGC 5128      & 13:25:27.6152 & -43:01:08.806 & 2      & 0.55     & 3.8  & 1.5      & 0.6  & 39 & N, B\\

IC 4296	& & 13:36:39.0325	&-33:57:57.073	& 0.185	& 13.5 &	50.8&	2.7&	1.1 & 7, 12& N, B, L\\

NGC 5252	&& 13:38:15.9633 &	+04:32:33.294	& 0.0105 &	9.5	& 92	& 1.1	& 0.4 & 7, 40& L\\

PGC 049940 & &	14:01:41.8426	& -11:36:24.968 &	0.0431 &	36.1	& 147.2 &	2.5 &	1.0 & 7, 41& L\\

NGC 5846	& &15:06:29.2320	&+01:36:22.644	&0.0138	&11&	24.9&	4.5	& 1.8 & 7, 12& L\\ 

3C 317 & UGC 9799      & 15:16:44.5070    & +07:01:18.078    & 0.034   & 45.7      & 145   & 3.2      & 1.3  & 10, 42& N, Z, B\\

Mrk 501     &UGC 10599         & 16:53:52.2166    & +39:45:36.608    & 0.279   & 34  & 147    & 2.4      & 1.0  & 10, 43& N\\

Cyg A*    &3C 405           & 19:59:28.3560    & +40:44:02.096    & 0.8   & 25     & 244.7 & 1.0      & 0.4  & 44, 45 &N, L\\

IC 1459   &           & 22:57:10.6070    & -36:27:43.996    & 0.2067    & 26      & 20.5   & 13.0     & 5.2  & 7, 46, 47 &N, Z, B, L\\ \bottomrule
\end{tabular}
\begin{tablenotes}
	\footnotesize {
\item\textbf{Notes:} Columns are as follows: (1) the name of candidates; (2) other names of candidates; (3) right ascension; (4) declination; (5) the observed or estimated flux density at 230\,GHz; (6) the mass of SMBH; (7) the distance between the SMBH and the Earth; (8) the angular diameter of emission ring (black hole shadow size), estimated by $\theta=2\sqrt{27}GM_{\rm BH}/(D_{\rm BH} c^2)$ \citep[see e.g.,][]{Falcke2013CQGra30.24.244003}; (9) the angular width of emission ring, estimated by $40\%$ of $\theta$ (consistent with the current EHT measurement of Sgr~A* and M87*); (10) references; (11) references from which the source was selected: N is \cite{Nair2024evn..conf...75N}, Z is \cite{Zhang2025ApJ985.41}, B is \cite{BenZineb2024arXiv2412.01904}, L is \cite{Lo2023ApJ950.10}. Candidates are arranged in order of their RA. Sgr A* and M87* are listed at the beginning for comparison, with the corresponding values from EHT results.
\item\textbf{References:}
    \tnote{1}\cite{EHTC2022ApJ930L13}.
    \tnote{2}\cite{EHTC2022ApJ930L12}.
    \tnote{3}\cite{GRAVITY2022AA657.12}.
    \tnote{4}\cite{EHTC2019ApJL875.L4}.
    \tnote{5}\cite{EHTC2019ApJL875.L1}.
    \tnote{6}\cite{EHTC2019ApJL875.L6}.
    \tnote{7}\cite{Lo2023ApJ950.10}.
    \tnote{8}\cite{Bender2005ApJ631.280}.
    \tnote{9}\cite{Stanek1998ApJ503.131}.
    \tnote{10}ALMA Calibrator Source Catalog (ACSC, \url{https://almascience.eso.org/sc/}) band 6 data.
    \tnote{11}\cite{Boizelle2021ApJ908.19}.
    \tnote{12}\cite{Sani2011MNRAS413.1479}.
    \tnote{13}\cite{Baczko2024AA692.205}.
    \tnote{14}\cite{Woo2002ApJ579.530}.
    \tnote{15}\cite{Kameno2020ApJ895.73}.    
    \tnote{16}\cite{Agudo2014AA566.59}, for NGC 4261 we use the observed $F_{\rm 86\,GHz}$ as a rough approximation to $F_{\rm\,230\\,GHz}$.
    \tnote{17} $M_{\rm BH}$ is calculated by the stellar velocity dispersion 341.8 km/s measured by \cite{vandenBosch2015ApJS218.10} and the $M_{\rm BH}-\sigma$ relation in the reference. 
    \tnote{18}$D_{\rm BH}$ is calculated from z=0.028302 \citep{vandenBosch2015ApJS218.10}
    \tnote{19}\cite{Nagai2019ApJ883.193}.
    \tnote{20}\cite{Riffel2020MNRAS496.4857}.
    \tnote{21}\cite{Walsh2016ApJ817.2}.
    \tnote{22}\cite{Houghton2006MNRAS367.2}.
    \tnote{23}\cite{Tonry2001ApJ546.681}.
    \tnote{24}\cite{Thomas2016Natur532.340}.
    \tnote{25} $F_{\rm 230\,GHz}$ is estimated by \cite{Zhang2025ApJ985.41}.
    \tnote{26}\cite{Gultekin2011ApJ738.17}.
    \tnote{27}$F_{\rm 230\,GHz}$ is estimated by \cite{Ramakrishnan2023Galax11.15}.
    \tnote{28}\cite{Devereux2003AJ125.1226}.
    \tnote{29}\cite{Bartel2007ApJ668.924}.
    \tnote{30}\cite{Anton2004MNRAS352.673}.
    \tnote{31}\cite{Balasubramaniam2021ApJ922.84}.  
    \tnote{32}\cite{Walsh2012ApJ753.79}.
    \tnote{33}\cite{Davis2013Natur494.328}.
    \tnote{34}We use the 96\,GHz flux density observed by \cite{Doi2011AJ142.167} as a rough approximation to $F_{\rm\,230\,GHz}$; $M_{\rm BH}$ and $D_{\rm BH}$ are also from the reference.
    \tnote{35}\cite{Menezes2015ApJ808.27}.
    \tnote{36}\cite{McQuinn2016AJ152.144}.
    \tnote{37}\cite{Osorno2025AA695.72O}.
    \tnote{38}\cite{Chen2023ApJ951.93}.
    \tnote{39}\cite{Janssen2021NatAs5.1017}.
    \tnote{40}\cite{Capetti2005AA431.465}.
    \tnote{41}\cite{DallaBonta2009ApJ690.537}.
    \tnote{42}\cite{Mezcua2018MNRAS474.1342}.
    \tnote{43}\cite{Barth2002ApJ566.L13}.
    \tnote{44}\cite{Lo2021ApJ911.35}.
    \tnote{45}\cite{Tadhunter2003MNRAS342.861}.
    \tnote{46}\cite{Cappellari2002ApJ578.787}.
    \tnote{47}\cite{Tingay2015MNRAS448.252}    
    }
\end{tablenotes}
\label{tab:BHimages_candidates}
\end{threeparttable}
\end{table*}

\begin{figure*}
    \centering
    \includegraphics[width=0.9\linewidth]{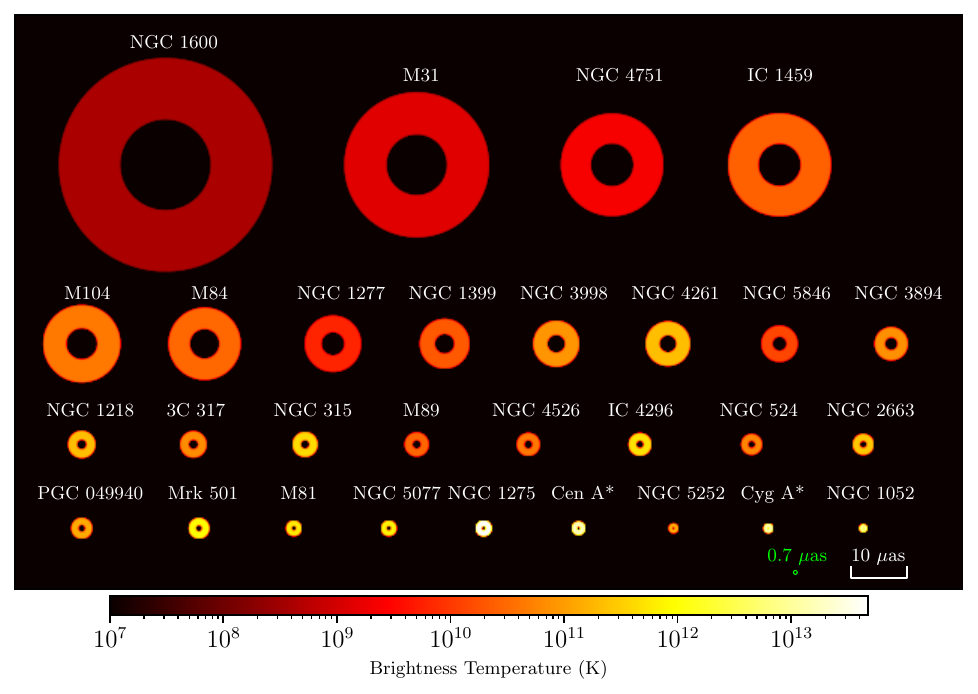}
    \caption{Ring model images of 29 SMBH candidates (Table~\ref{tab:BHimages_candidates} excludes Sgr~A* and M87*), ordered by their ring sizes. All images are shown with a common angular scale and color map. The green circle in the lower right indicates a 0.7~$\mu$as beam estimated from Eq.~\eqref{eq:theta_beam}. 
 }
    \label{fig:model_image}
\end{figure*}

A uniform brightness distribution ring is used to represent the basic morphology of a candidate SMBH, and its diameter and width are predicted by the mass and distance. Fig.~\ref{fig:model_image} displays the geometric ring model images of 29 candidates (excluding the already captured M87* and Sgr A*), using the identical color scaling and angular scaling to facilitate comparison of intensities and ring sizes.

\subsection{Geometric model visibilities}
\label{subsec:geom_model}

VLBI directly observes visibilities rather than images; accordingly, our observational simulations and detectability analysis are performed entirely in the visibility domain. For the basic shadow detectability analysis (Section~\ref{sec:detectability} excluding~\ref{sec:simobs_M104}), each candidate source is modeled with a simple geometric ring. This approach provides a clear physical interpretation and enables straightforward, quantitative criteria for shadow detection. Beyond this, an azimuthal asymmetry ring model is used for estimating asymmetry uncertainties in Appendix~\ref{sec:asymmetry}, and a general relativistic magnetohydrodynamic (GRMHD) model for the simulated observational case study in Section~\ref{sec:simobs_M104}.

For a geometric ring model with total flux density $V_0$, diameter $\hat{d}$ and width $\hat{w}$, its visibility observed in the ($u,v$) coordinates can be calculated by \citep{Kamruddin2013MNRAS434.765}:
\begin{equation}
    V(u,v;V_0,\hat{d},\hat{w})=
    \frac{2V_0}{k(a_2^2-a_1^2)}\left[a_2J_1(ka_2)-a_1J_1(ka_1)\right],
    \label{eq:ring_model_vis}
\end{equation}
where
\begin{equation}
    k=\frac{2\pi\sqrt{u^2+v^2}}{\lambda},\quad
    a_1=\dfrac{\hat{d}-\hat{w}}{2},\quad
    a_2=\dfrac{\hat{d}+\hat{w}}{2},
\end{equation}
$a_1$ and $a_2$ are the inner and outer radii, respectively. $J_1$ denotes the Bessel function of the first kind.

In the visibility domain, a defining feature of the ring model is that its visibility amplitude decreases with baseline ($\sqrt{(u^2+v^2)}/\lambda$) in a damped oscillatory manner, with local minima corresponding to nulls (zero points). This behavior is quantitatively distinct from that of a Gaussian model, whose visibility amplitude decreases monotonically. Among the parameters of the ring model, the ring diameter $\hat{d}$ plays the primary role in governing the visibility structure, specifically in determining the baseline at which the first null occurs. The ring width $\hat{w}$ plays a secondary role by mainly modulating the amplitudes of the oscillations. Varying the fractional width within a plausible range of 30\%-50\% \citep[e.g., the EHT ring width measurement for Sgr A*;][]{EHTC2022ApJ930L12} leads to only modest changes in the secondary peak amplitude ($\sim16\%$), while leaving the null locations essentially unchanged. Comparisons of visibility amplitudes for different geometric models and ring parameters are illustrated in Figures 10.11 and Fig. 10.12 in \cite{Thompson2017isra.book}. 

\section{Lunar-based telescope assumptions}\label{sec:lunar_telescope}

\subsection{Site candidates}\label{sec:sites}

\begin{figure}
    \centering
    \includegraphics[width=0.95\linewidth]{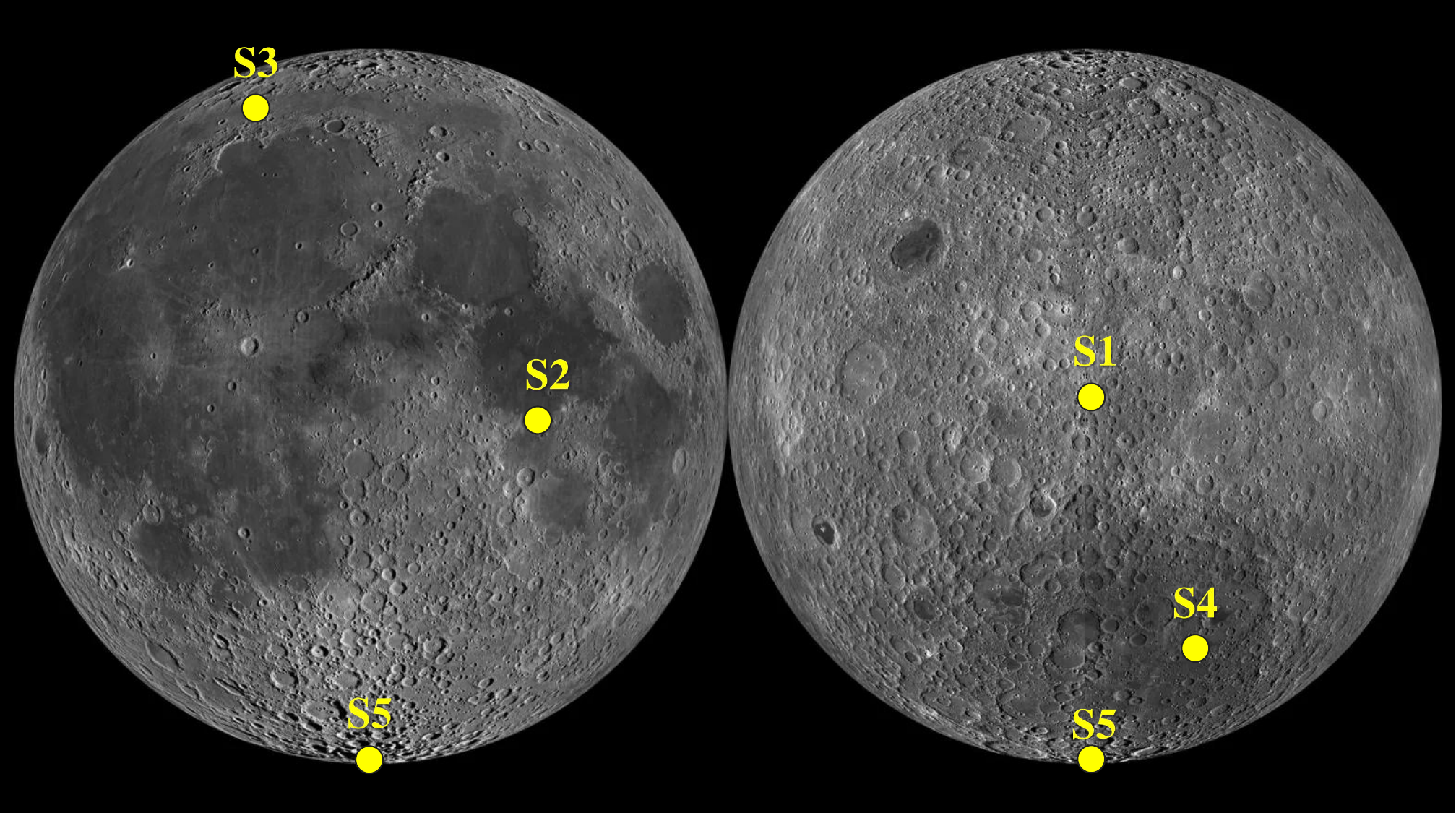}
    \caption{Map of telescope sites on the Moon.}
    \label{fig:sites}
\end{figure}

We select five candidate sites for the lunar-based telescope (Fig.~\ref{fig:sites}, Table~\ref{tab:sites}), chosen for their relevance to both baseline and historic or proposed lunar missions. These include S1, the point farthest from Earth, providing the maximum possible Moon–Earth baseline; the historic landing sites S2 (Apollo 11), S3 (Chang’E-5), and S4 (Chang’E-6); and S5, the lunar south pole, a leading candidate location for the ILRS mission. All site coordinates are given in the Mean Earth coordinate system, with differences from the Principal Axis system neglected \citep[$\sim$860 m;][]{Wang2021RemS13.590}.

\begin{table}
\begin{threeparttable}[b]
\centering
\caption{Coordinates of telescope sites on the Moon}
\renewcommand\arraystretch{1.3} 
\setlength\tabcolsep{8pt}       
\begin{tabular}{@{}cccccccc@{}}
\toprule
Site & Longitude & Latitude & Description\\ \midrule
S1 & $180^\circ$ E & $0^\circ$ & Farthest point from Earth\\
S2 & $23.47^\circ$ E & $0.67^\circ$ N & Apollo 11 site \tnote{(1)} \\
S3 & $51.92^\circ$ W & $43.06^\circ$ N & Chang'E-5 site \tnote{(2)}\\
S4 & $153.98^\circ$ W & $41.63^\circ$ S & Chang'E-6 site \tnote{(3)}\\
S5 & $0^\circ$ E & $90^\circ$ S & South pole\\
\bottomrule
\end{tabular}
\begin{tablenotes}
	\footnotesize {
        \item [(1)] \cite{Davies2000JGR105.20277}.
        \tnote{(2)}\cite{Wang2021RemS13.590}.
        \tnote{(3)}\cite{Liu2024NRSB28.6.1648}
    }
\end{tablenotes}
\label{tab:sites}
\end{threeparttable}
\end{table}

We assume the lunar-based telescope can observe a target only when its elevation angle exceeds 15$^\circ$. This elevation threshold is adopted to account for potential obscuration by the local lunar surface topography and is consistent with a commonly used minimum elevation constraint for ground-based radio telescopes.
Unlike ground-based telescopes, a lunar-based telescope’s source elevation varies periodically with the Moon’s sidereal rotation. We calculate the total observable duration (in days) for each site over one lunar sidereal month for any celestial location.

\begin{figure*}
    \centering
    \includegraphics[width=0.98\linewidth]{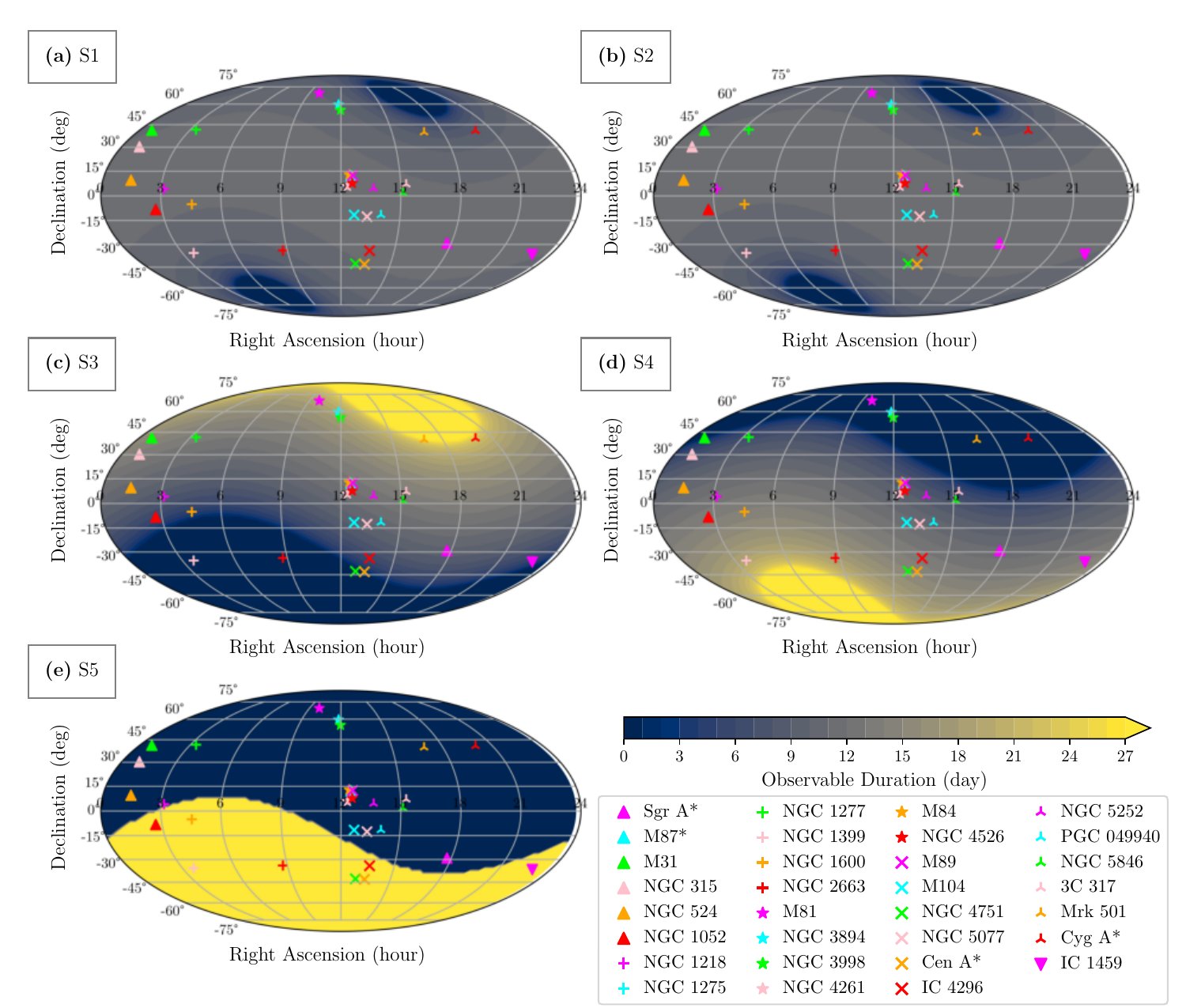}
    \caption{Observable duration maps for different sites over one lunar sidereal month (27.32 days), calculated by accumulating the observable time when the elevation exceeds 15$^\circ$ for a source located at any right ascension and declination on the entire celestial sphere.}
    \label{fig:obs_time}
\end{figure*}

Figure~\ref{fig:obs_time} presents the resulting sky maps of observable durations for sites S1-S5 (panels a-e), with the positions of SMBH candidates overlaid. Each map is constructed by evaluating the elevation angle of sources located on a grid across the celestial sphere and accumulating the time over one sidereal month during which the elevation exceeds 15$^\circ$. This accumulated time is defined as the observable duration and is represented by color bars. We compare the differences in map patterns across different sites and assess their impact on all candidate sources. 

The results show a strong latitudinal dependence of the observable duration. Near-equatorial sites (S1, S2) provide an approximately 11-day observable window for most candidates. Site S3 offers longer observable time for northern sky sources (e.g., M81, NGC~3894, NGC~3899, Mrk~501, Cyg~A*), while Site S4 provides longer durations for southern sources (e.g., NGC~1399, NGC~2663, NGC~4751, Cen~A*). Site S5 exhibits a sharp hemispheric dichotomy: southern sources remain continuously observable for a full month, whereas northern sources are entirely unobservable. This observable-unobservable transition follows a sinusoidal pattern resulting from the inclination between the lunar orbital plane and the celestial equator. Consequently, a telescope at S5 can observe only eight candidates: NGC~1052, NGC~1399, NGC~1600, NGC~2663, NGC~4751, Cen~A*, IC~4296, and IC~1459.

Site S1 is selected as the location for the lunar-based telescope in all the following calculations and analyses due to its capability to provide an observable time window for all candidate sources.

\subsection{Antenna size and baseline sensitivity}
\label{sec:sensitivity}

The performance of a lunar-based 230\,GHz telescope depends on several factors, including antenna size, aperture efficiency, system temperature, and bandwidth, all of which directly affect baseline sensitivity and, consequently, the detectability of  SMBH shadows. In this work, we adopt antenna size as the primary variable, as it is typically the most practically constrained parameter in mission considerations, while the other parameters are fixed to representative values. Parameterizing baseline sensitivity in terms of antenna size provides a direct and intuitive measure of the practical effort required to achieve the detectability for each SMBH shadow candidate.

For a telescope with system temperature $T_{\rm sys}$ and effective area $A_{\rm eff}$, the System Equivalent Flux Density (SEFD) is defined as 
\begin{equation}
    \rm{SEFD}=2k_{\rm B}\frac{T_{\rm sys}}{A_{\rm eff}},
    \label{eq:SEFD}
\end{equation} 
where $k_{\rm B}$ is the Boltzmann constant. As a reference, ALMA (equivalent diameter 73 m, aperture efficiency 0.68, $T_{\rm sys}=76$ K) has an SEFD of 74 Jy at 230\,GHz , adopted from the reported value for the EHT 2017 observation at median elevations and typical atmospheric opacities \citep{EHTC2019ApJL875.L2}. We keep this value constant across elevations as a representative reference, while noting that in practice the SEFD varies with elevation and atmospheric conditions.
Assuming a lunar-based telescope with $T_{\rm sys}=100$ K and 50\% aperture efficiency, SEFD for various antenna sizes are calculated by Eq.~\ref{eq:SEFD} and summarized in Table~\ref{tab:sensitivity}. 

The minimum detectable flux density in VLBI is constrained by the baseline sensitivity, which can be estimated from SEFD. For a baseline formed by telescope i and telescope j, the standard deviation of thermal noise is 
 \citep{Thompson2017isra.book}
\begin{equation}
    \sigma_{\rm ij}=\frac{1}{\eta}\sqrt{\frac{\rm SEFD_i\times SEFD_j}{2\Delta\nu\Delta t}},
    \label{eq:thermal_noise}
\end{equation}
where $\eta$ is the digital loss factor, $\Delta\nu$ is the bandwidth and $\Delta t$ is the integration time. We estimate the thermal noise of the lunar-based telescope and ALMA baseline $\sigma_{\rm LT-ALMA}$ with the following parameters: $\eta=0.88$ (2-bit sampling), $\Delta\nu=32$\,GHz, $\Delta t=10$ min. The adopted bandwidth and integration time are consistent with those proposed for discussing using Moon–Earth VLBI detecting higher-order photon rings in \cite{Johnson2020SciA6.1310}. While the current bandwidth of the EHT is 8 GHz \citep{EHTC2019ApJL875.L2}, increasing bandwidth is a key goal of the ngEHT \citep{Doeleman2019BAAS51.256,Johnson2023Galax11.61}, making 32 GHz a plausible near-future capability. Regarding integration time, atmospheric phase fluctuations typically limit integration times for ground-based millimeter VLBI to $\sim$ 10 s, but frequency phase transfer technique \citep{Rioja2011AJ141.114,Rioja2023Galax11.16} can extend this timescale to several hours \citep{Jiang2023ApJ959.11}. Our choice of 10 min is comparable to a standard scan for EHT observation. In practice, achieving such an integration time for high-frequency, long-baseline VLBI requires dedicated fringe-fitting strategies, and additional effects such as uv-smearing should also be taken into consideration. Developing such fringe-fitting strategies is an important direction for future work and is beyond the scope of this work. 
Finally, by considering 7$\sigma$ detection threshold, we use $7\sigma_{\rm LT-ALMA}$ as an indicator for Moon-Earth baseline sensitivity. Thus, for a lunar-based telescopes with antenna diameters of 5, 10, 20, and 40 m, the corresponding Moon–Earth baseline sensitivities are 1.852, 0.926, 0.463, and 0.232 mJy, details are reported in Table~\ref{tab:sensitivity}.   

\begin{table}
\begin{threeparttable}[b]
\centering
\caption{Baseline sensitivity of different antenna diameters}
\label{tab:sensitivity}
\renewcommand\arraystretch{1.3} 
\setlength\tabcolsep{18pt}       
\begin{tabular}{@{}cccccccc@{}}
\toprule
Diameter  & SEFD\tnote{(1)}  & $\sigma_{\rm LT-ALMA}$\tnote{(2)} & Sensit.\tnote{(3)} \\ 
(m) &(Jy)  &(mJy)  &(mJy) \\ 
\midrule
5 & 28126 &  0.265 & 1.852\\
10 & 7032 &  0.132 & 0.926\\
20 & 1758 &  0.066 & 0.463\\
40 &  440 &   0.033 & 0.232\\
\bottomrule
\end{tabular}
\begin{tablenotes}
	\footnotesize {
        \item [(1)] SEFD of the lunar-based telescope is calculated by Eq.~\eqref{eq:SEFD}, with the assumed aperture efficiency of 0.5, $T_{\rm sys}=100$ K. \\
        \tnote{(2)}$\sigma_{\rm LT-ALMA}$ is the thermal noise standard deviation for the baseline of lunar-based telescope and ALMA, calculated by Eq.~\ref{eq:thermal_noise} with SEFD$_{\rm ALMA}=74$ Jy, $\eta=0.88$, $\Delta\nu=32$\,GHz, $\Delta t=10$ min at the observing frequency 230\,GHz.\\
        \tnote{(3)} Baseline sensitivity is calculated by $7\sigma_{\rm LT-ALMA}$ detection threshold.
    }
\end{tablenotes}
\end{threeparttable}
\end{table}

\section{Moon-Earth baseline coverage simulation}\label{sec:VLBIsimulation}

Simulating VLBI baseline coverage requires computing the time-dependent relative positions of all array elements and the target source. The inclusion of a lunar-based telescope introduces additional complexity due to the need for precise coordinate transformations. We address this using Omni\textit{UV}, a public toolkit designed for multipurpose simulations of space and ground VLBI observations\footnote{\url{https://github.com/liulei/omniuv}}\citep{Liu2022AJ164.67}. 

The Omni\textit{UV} toolkit incorporates lunar-based stations into VLBI simulations by precisely computing its station trajectory. The lunar station position is transformed from a Lunar Fixed Frame to a Lunar Celestial Frame (LCF) using the time-dependent rotation matrices derived from the JPL Planetary Ephemeris (version DE421) Lunar PCKs (Planetary Constants Kernels). The LCF coordinates are then converted to the International Celestial Reference Frame (ICRF) using Moon-Earth vectors from the JPL Planetary Ephemeris (version DE421) SPK (Spacecraft and Planet Kernel). 
Earth-based station coordinates are transformed from the International Terrestrial Reference Frame (ITRF) to the ICRF following the International Earth Rotation and Reference Systems Service (IERS) Conventions (2010) \citep{2010ITN....36....1P}, with the Earth Rotation Angle (ERA) treated as the primary component. With all station trajectories referenced to the ICRF, ($u,v$) coverage is computed following the definition and formalism of \cite{Thompson2017isra.book}, where ($u,v$) plane is perpendicular to the source direction and the u-axis aligned to the celestial east direction.  

We simulated ($u,v$) coverage for all SMBH candidates assuming an array combines the ground-based EHT with a lunar-based telescope at site S1 (see Section~\ref{sec:sites}). The simulation includes the following constraints: 1) a minimal elevation angle of 15$^\circ$ for both lunar-based and ground-based telescopes; 2) a minimal angular separation threshold of 5$^\circ$ between the source and both the solar and lunar limbs; 3) continuous observations over one lunar sidereal month. An entire lunar sidereal month is chosen to capture the full Moon–Earth baseline variations and maximize ($u,v$) plane sampling. 

\begin{figure*}
    \centering
    \includegraphics[width=0.98\linewidth]{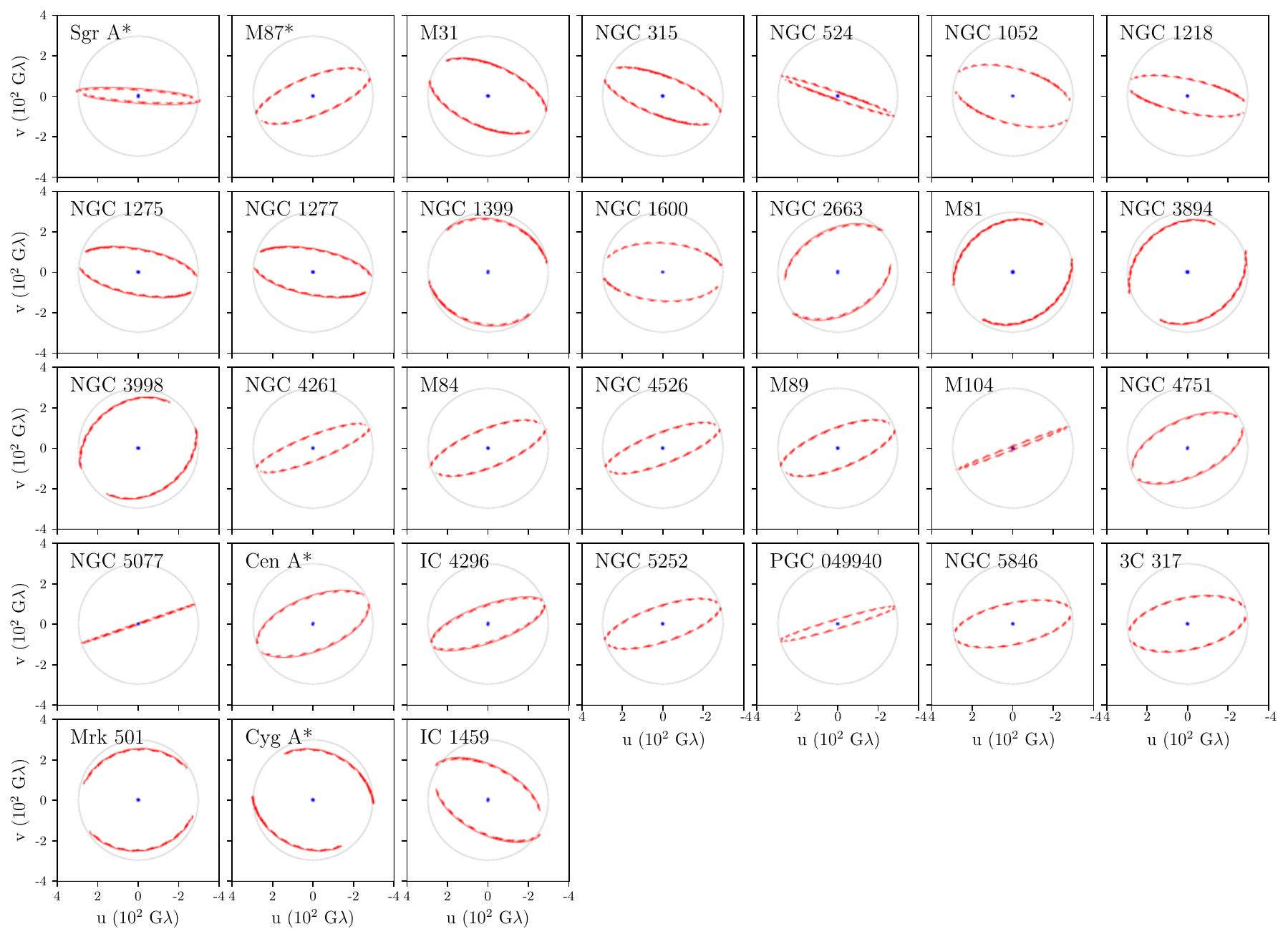}
    \caption{Simulated ($u,v$) coverage for a Moon-Earth VLBI array, formed by a lunar-based telescope at S1 and the EHT 2025 array. Data points for the Moon-Earth baselines are shown in red, and for the ground-based EHT baselines in blue. The grey circle indicates a baseline of 30 $D_\oplus$.}
    \label{fig:uv_coverage}
\end{figure*}

The results (Figure~\ref{fig:uv_coverage}) show dual arc-shaped coverage patterns for Moon-Earth baselines (red dots), separated from ground-only baselines (blue) by a coverage gap due to the absence of intermediate telescopes. The morphology of the Moon-Earth baselines varies with the source position, from linear to elliptical to circular. It depends on the geometric relationship between the source direction and the Moon-Earth baseline direction. Specifically, linear patterns occur when the two direction are in the coplanar case, the circular pattern corresponds to the orthogonal case, and the elliptical pattern manifests in the intermediate case. 

The minimal projected Moon-Earth baseline in the ($u,v$) coverage is a very important quantity for comparison with the resolvability criteria in Section~\ref{sec:detectability}. The dependence of the minimal projected Moon–Earth baseline on source location follows a sinusoidal pattern on the sky map, reflecting the geometry of the Moon's orbit (5.15$^\circ$ mean inclination relative to the Earth's equator); see Appendix~\ref{sec:minimal_baseline} and Figure~\ref{fig:BLmin} for details.

\section{Detectability analysis}\label{sec:detectability}
\subsection{Criteria of detection}\label{sec:criteria}

Experience from ground-based 1.3 mm VLBI imaging of M87* and Sgr~A* shows that the first step in detecting a black hole shadow is to distinguish the surrounding emission ring from an unresolved Gaussian component. For sources beyond M87* and Sgr~A*, the extremely sparse ($u,v$) coverage provided by Moon–Earth baselines (Fig.~\ref{fig:uv_coverage}) is insufficient for reliable image reconstruction. We therefore assess the detectability of the ring structure directly in the visibility domain. A critical discriminator between ring and Gaussian geometric models in this domain is the presence of nulls in the visibility amplitude profile (see Section~\ref{subsec:geom_model}).
We therefore focus on detecting the first null in the visibility amplitudes as the primary indicator of a black hole shadow. When the range of projected Moon–Earth baselines spans the baseline length corresponding to this first null, the visibility data around the null can be sampled, allowing its position to be robustly determined. For a given SMBH candidate, the resolvability of the shadow can thus be evaluated by comparing the predicted location of the first null with the projected Moon–Earth baseline range. 

In addition, VLBI sensitivity imposes a further constraint on detecting the first null. A practical criterion is that 10\% of the secondary peak amplitude should remain above the baseline sensitivity threshold \citep[consistent with EHT observations of Sgr~A* and M87*; ][]{EHTC2019ApJL875.L1,EHTC2022ApJ930L12}. If this condition is not met, the post‑null visibility rise may become insufficient to trace the first null location, making it difficult to detect even if the null lies within the baseline range.

Although the primary ring structures of M87* and Sgr~A* have already been detected with ground-based VLBI, we apply the same criteria to these sources for consistency. It is of particular interest whether Moon–Earth baselines can resolve their photon rings, as we discuss in Section~\ref{sec:photon_ring}.

Based on these considerations, we evaluated the detectability of 31 candidate sources for Moon-Earth VLBI observations using the following two criteria:
\begin{itemize}
    \item \textbf{Within the resolvable region}: The first null location must lie within the projected Moon-Earth baseline range;
    \item \textbf{Above the sensitivity threshold}: 10\% of the secondary peak amplitude must exceed 0.232 mJy, corresponding to the baseline sensitivity of a 40 m lunar-based telescope paired with ALMA on the Earth.
\end{itemize}

\begin{figure*}
    \centering
    \includegraphics[width=0.95\linewidth]{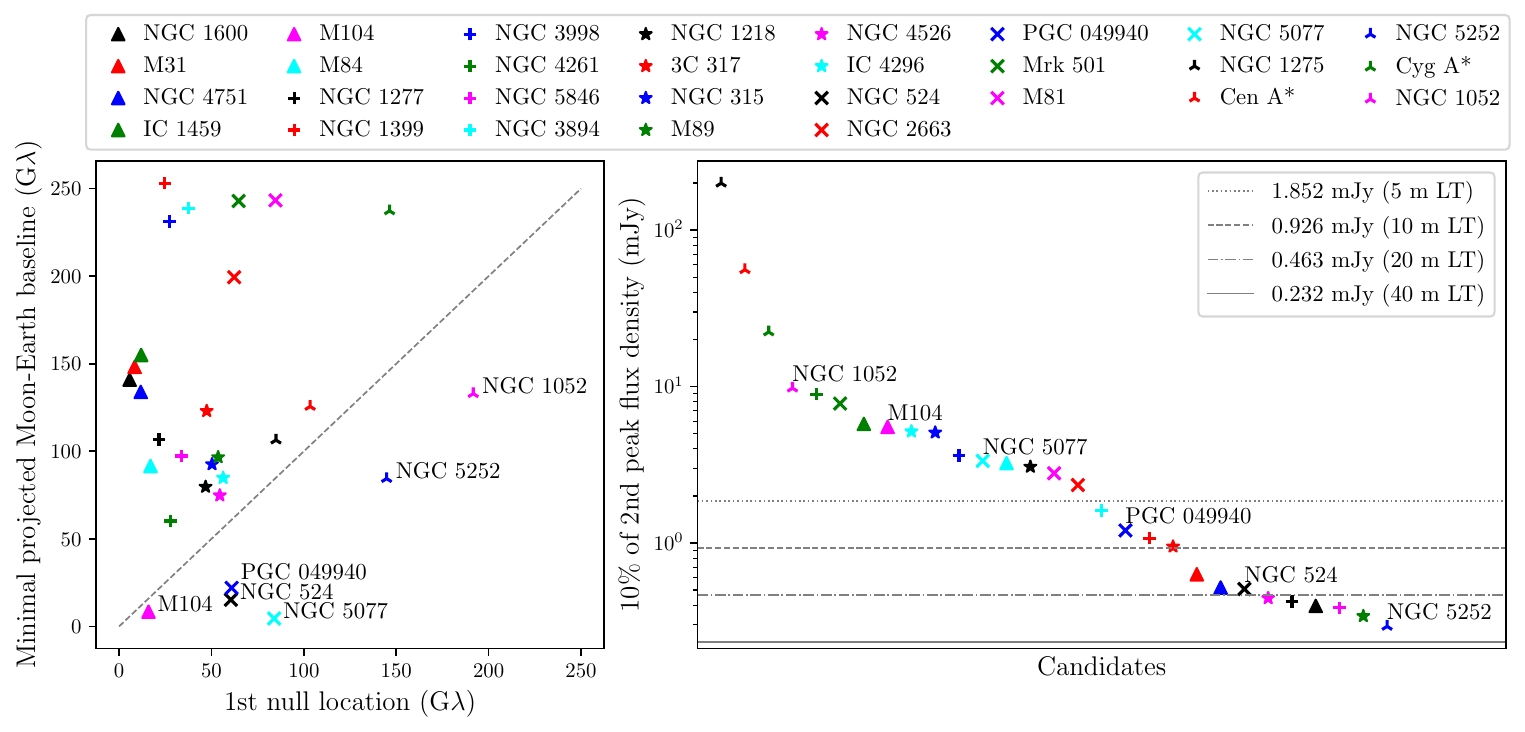}
    \caption{Left panel: First null location versus minimal projected Moon-Earth baseline; the grey dashed line marks equality. Sources below it are shadow-resolvable with Moon-Earth baselines: M104, NGC~524, PGC~049940, NGC~5077, NGC~5252, and NGC~1052. Right panel: 10\% of the secondary peak flux density compared against baseline sensitivity thresholds for ALMA and lunar-based telescopes (LT) of diameters 5, 10, 20, and 40 m (see Table~\ref{tab:sensitivity}).}
    \label{fig:meet_criteria}
\end{figure*}

The two criteria were applied to the 31 SMBH candidates, as shown in Figure~\ref{fig:meet_criteria}. The left panel illustrates the resolvability criterion: only six sources lie below the grey dashed line, indicating that their first null location exceeds the minimal projected Moon-Earth baseline. Notably, under this criterion, neither M87* nor Sgr~A* falls within the resolvable region, because the minimal projected Moon-Earth baselines are too long to resolve their first nulls. 
We compare the first null location only with the minimum projected baseline, not the maximum, because the candidate selection already ensures that the shadow size is larger than $\theta_{\rm beam}$, which guarantees that the first null location remains smaller than the maximum baseline. Hence, we do not need to compare the first null with the maximum projected baseline. The right panel in Figure~\ref{fig:meet_criteria} presents the sensitivity criterion: all 31 candidates exceed the 0.232 mJy threshold corresponding to a 40 m lunar-based telescope, while 16 of them remain detectable for a 5 m lunar-based telescope (1.852 mJy). For the 40 m case, all sources satisfy the sensitivity requirement.

\begin{figure*}
    \centering
    \includegraphics[width=0.95\linewidth]{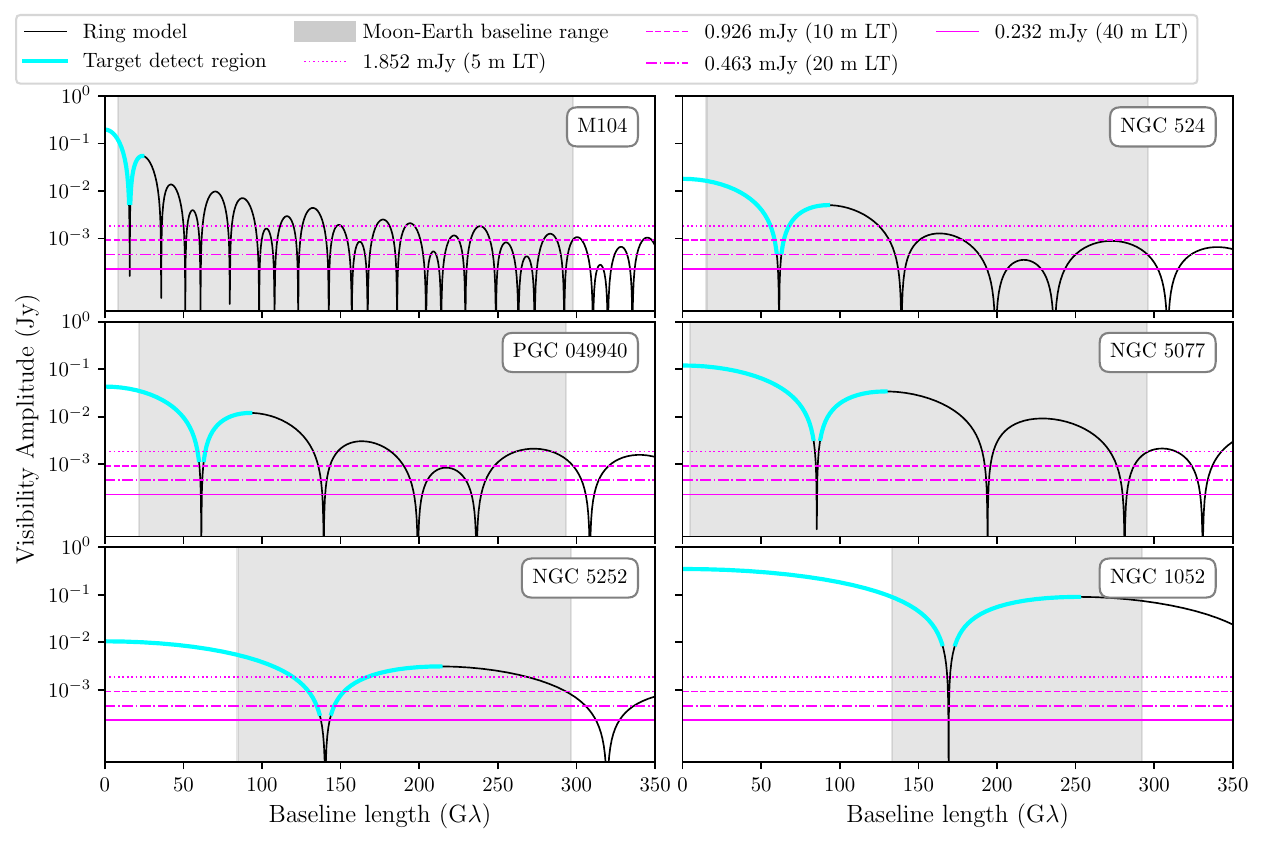}
    \caption{Ring-model visibility amplitudes (black solid lines) for the six Moon-Earth VLBI shadow-detectable candidates (see Tab.~\ref{tab:BHimages_candidates} and Fig.~\ref{fig:model_image} for ring model parameters and images). Cyan lines mark the region around the first null where the amplitude exceeds 10\% of the secondary peak. Grey shaded regions indicate the projected Moon-Earth baseline range for each source (M104: 8.4 to 298.1 G$\lambda$; NGC 524: 15.2 to 295.9 G$\lambda$; PGC 049940: 22.0 to 293.2 G$\lambda$; NGC 5077: 4.5 to 295.6 G$\lambda$; NGC 1052: 132.9 to 292.0 G$\lambda$; NGC 5252: 84.4 to 296.5 G$\lambda$). Horizontal lines denote baseline sensitivity for ALMA and lunar-based telescopes of diameters 5, 10, 20, and 40 m (see Table~\ref{tab:sensitivity}).}
    \label{fig:visibility}
\end{figure*}

Considering both criteria, M104, NGC 524, PGC 049940, NGC 5077, NGC 5252, and NGC 1052 qualify as unique shadow-detectable candidates for Moon-Earth VLBI. Figure~\ref{fig:visibility} shows their predicted ring-model visibility amplitudes, where the first nulls and regions exceeding 10\% of the secondary peak (cyan lines) can be directly compared with the observable baseline range (grey shaded region) and the sensitivity thresholds (magenta lines). These candidates are discussed individually in the following section.

\subsection{Shadow-detectable candidates}\label{sec:best}
\begin{itemize}
\item \textbf{M104}, also known as NGC 4594, is the SMBH in the center of the Sombrero Galaxy. The galaxy is an early-type spiral galaxy located in the Virgo cluster, classified as a Low Ionization Nuclear Emission-line Region \citep[LINER;][]{Heckman1980AA87.152}, with a viewing angle of $66^\circ$ \citep{Yan2024ApJ965.128}. We adopt a black hole mass of $9\pm2\times 10^8\ \rm{M}_\odot$ from stellar kinematics \citep{Menezes2015ApJ808.27}, a distance of $9.55\pm0.31$ Mpc \citep{McQuinn2016AJ152.144} and a 230\,GHz flux density of 0.198$\pm$0.01 Jy from ALMA band 6 data. A smaller mass of $6.6\pm0.4\times 10^8\ \rm{M}_\odot$ was measured by \cite{Jardel2011ApJ739.21} (distance is assumed to be 9.8 Mpc), and a similar value of $6.65^{+0.40}_{-0.41}\times 10^8\ \rm{M}_\odot$ from the fundamental plane was estimated by \cite{Gultekin2019ApJ871.80} (with a distance of 9.87 Mpc). M104 is an excellent target for high-resolution VLBI, including the next-generation EHT, BHEX and other space VLBI proposals \citep{Zhang2025ApJ985.41,Johnson2024SPIE13092E..2DJ,Ford2025ApJ981.126,Fish2020AdSpR65.821}, and is particularly well suited for Moon-Earth VLBI. Its large shadow size ($\sim 10\ \mu$as) and the linear Moon-Earth ($u,v$) coverage produce numerous nulls including the first null in the visibility amplitudes within the observable range. A key advantage is that the first null occurs at a relatively short baseline (15.8\,G$\lambda$, 1.6\,$D_\oplus$), facilitating fringe detection. Furthermore, the flux density of the 10\% secondary peak of the visibility amplitude is sufficiently high (5.51 mJy) to be detectable even with a 5 m lunar-based antenna. We additionally verify such detection with simulated visibility amplitude data generated from the GRMHD model (see Section~\ref{sec:simobs_M104}).

\item \textbf{NGC 524} is an early-type galaxy hosting a compact nucleus that is bright at low frequencies but faint at high frequencies. Its flux density is only 18.2$\pm$2.8 mJy at 228.9\,GHz \citep[][adopted in this work]{Lo2023ApJ950.10} and 8.3$\pm$0.1 mJy at 237.3\,GHz \citep{Smith2019MNRAS485.4359}. Black hole mass and distance measurements are varies: $8.32^{+0.60}_{-0.37}\times 10^8\ \rm{M}_\odot$ and 33.6 Mpc in \cite{Sani2011MNRAS413.1479} (adopted in this work); $8.67^{+0.94}_{-0.46}\times 10^8\ \rm{M}_\odot$ and 24.22 Mpc in \cite{Gultekin2019ApJ871.80}; $8.3^{+2.7}_{-1.3}\times 10^8\ \rm{M}_\odot$ in \cite{Krajnovic2009MNRAS399.1839} and $4.0^{+3.5}_{-2.0}\times 10^8\ \rm{M}_\odot$ in \cite{Smith2019MNRAS485.4359}, with the same distance 23.3$\pm$2.3 Mpc \citep{Tonry2001ApJ546.681}. NGC 524 is one of the best candidates because of its linear-shaped ($u,v$) coverage, which can cover the first null (61.3\,G$\lambda$, 6.3\,$D_\oplus$) in the visibility amplitudes. The flux density is relatively low compared to other candidates. Its 10\% secondary peak is about 0.508 mJy, requiring the sensitivity of a 20 m antenna on the Moon.

\item \textbf{PGC 049940} is located in the Brightest Cluster Galaxy (BCG) Abell 1836 (A1836). We adopt a black hole mass of $3.61^{+0.41}_{-0.50}\times 10^9\ \rm{M}_\odot$ and a distance of 147.2 Mpc, estimated from gas kinematics \citep{DallaBonta2009ApJ690.537}. The fundamental plane relation suggests a mass of $3.9^{+0.4}_{-0.6}\times 10^9\ \rm{M}_\odot$ at 157.5 Mpc \citep{McConnell2011Natur480.215}. The source is weakly active and underluminous in radio \citep{Stawarz2014ApJ794.164}. It is observed to range from 29.1 to 43.1 mJy at 222\,GHz, with an error of $\sim$4 mJy \citep{Lo2023ApJ950.10}. Our results show the first null of visibility amplitudes is located at (61.3\,G$\lambda$, 6.3\,$D_\oplus$) and the 10\% secondary peak is about 1.20 mJy, which needs a lunar-based telescope with size larger than 10 m.  

\item \textbf{NGC 5077} is an elliptical galaxy. We adopt a black hole mass of 
$8.0^{+5.0}_{-3.3}\times 10^8\ \rm{M}_\odot$ and a distance of 44.9 Mpc \citep{Sani2011MNRAS413.1479}. Alternative estimates include
$6.8^{+4.3}_{-2.8}\times 10^8\ \rm{M}_\odot$ and 38 Mpc\citep{deFrancesco2008AA479.355},
$8.55^{+4.35}_{-4.48}\times 10^8\ \rm{M}_\odot$ and 38.7 Mpc \citep{Gultekin2019ApJ871.80}. At 230\,GHz, Submillimeter Array (SMA) monitoring from 2015 to 2019 yields a mean flux density of $0.120\pm0.033$ Jy \cite{Chen2023ApJ951.93}. The predicted first null in the visibility amplitude occurs at 85.4\,G$\lambda$ (8.7\,$D_\oplus$), with a 10\% secondary peak of 3.41 mJy, above the sensitivity threshold for the 5 m lunar-based telescope case. The highly linear ($u,v$) coverage for NGC 5077 enables sampling of its visibility profile over a wide range of baseline.

\item \textbf{NGC 5252} is an early-type (S0) Seyfert 2 galaxy at a distance of 92 Mpc, hosting a SMBH of mass $0.95^{+1.45}_{-0.45}\times 10^9\ \rm{M}_\odot$ \citep{Capetti2005AA431.465}. VLBI imaging at 1.7\,GHz suggests the presence of a dual AGN system with a separation of $\sim$10 kpc \citep{Yang2017MNRAS464.70}. At 222\,GHz, SMA observed a flux density of 4.9-10.5 mJy with errors about 2 mJy, which is speculated to originate from jet emission \citep{Lo2023ApJ950.10}. Although the ($u,v$) coverage of NGC 5252 is less linear than for M104, NGC 524, PGC 049940 and NGC 5077, NGC 5252 remains a detectable candidate for Moon-Earth VLBI due to its favorable visibility structure. The first null in the visibility amplitudes is expected at 140.2\,G$\lambda$ (14.3\,$D_\oplus$), which falls within the observable range (84.4-296.5\,G$\lambda$). The source's 10\% secondary peak is predicted to be weak (0.311 mJy), requiring a 40 m lunar-based telescope to meet the sensitivity requirements.

\item \textbf{NGC 1052} is a low-luminosity AGN, classified as a Seyfert 2 giant elliptical galaxy, hosting a SMBH of mass $1.54\times 10^8\ \rm{M}_\odot$ \citep{Woo2002ApJ579.530}. We adopt a distance of 17.6 Mpc \citep{Kameno2020ApJ895.73}, noting alternative estimates of 19.23 Mpc \citep{Tully2013AJ146.86} and 19.08 Mpc \citep{Lo2023ApJ950.10}. NGC 1052 is an EHT target with an observed compact flux density of 0.35$\pm$0.05 Jy \citep{Baczko2024AA692.205}, slightly lower than the SMA-only observation of 0.4965$\pm$0.0498 Jy \citep{Lo2023ApJ950.10}. Although the expected shadow size is small (only 0.9 $\mu$as), the Moon-Earth baseline range (132.9-292.0 G$\lambda$) spans the first null (169.1\,G$\lambda$, 17.3\,$D_\oplus$) in the visibility amplitudes. Owing to its high flux density, this ring structure is readily detectable even with a 5 m lunar-based telescope. 

\end{itemize}

This detectability depends on the mass and distance of the SMBH candidates. As discussed in Appendix~\ref{sec:uncertainties}, alternative values of these two parameters yield robust results for M104, NGC~524, PGC~049940, and NGC~5077, but less certain results for NGC~1052 and NGC~5252. Brightness distribution asymmetry introduces additional complexity (see Appendix~\ref{sec:asymmetry}, using M104 as an example), causing the first null to become "shallower", with the degree of shallowness depending on the peak brightness position angle.

\subsection{simulated visibility from GRMHD image model for M104}
\label{sec:simobs_M104}

To further verify that the proposed Moon–Earth VLBI configuration can robustly detect the first null and discriminate ring structures, we perform simulated observations using a GRMHD image model. We choose M104 as the representative case, which may also be applicable to other shadow-detectable candidates. The model is based on a non‑spin black hole Magnetically Arrested Disk (MAD) GRMHD simulation with an inclination of 163$^\circ$ \citep{Zhang2024AA687.88}, rescaled to match the expected shadow size (9.7 $\mu$as) and 230\,GHz flux density (0.198 Jy) of M104. From this image, we generate synthetic visibility amplitudes based on the simulated $(u, v)$ coverage (Section~\ref{sec:VLBIsimulation}) and add thermal noise assuming a 5 m lunar‑based telescope, using the tool \textit{ehtim}
\citep{Chael2018ApJ...857...23C}.

\begin{figure*}
    \centering
    \includegraphics[width=0.98\linewidth]{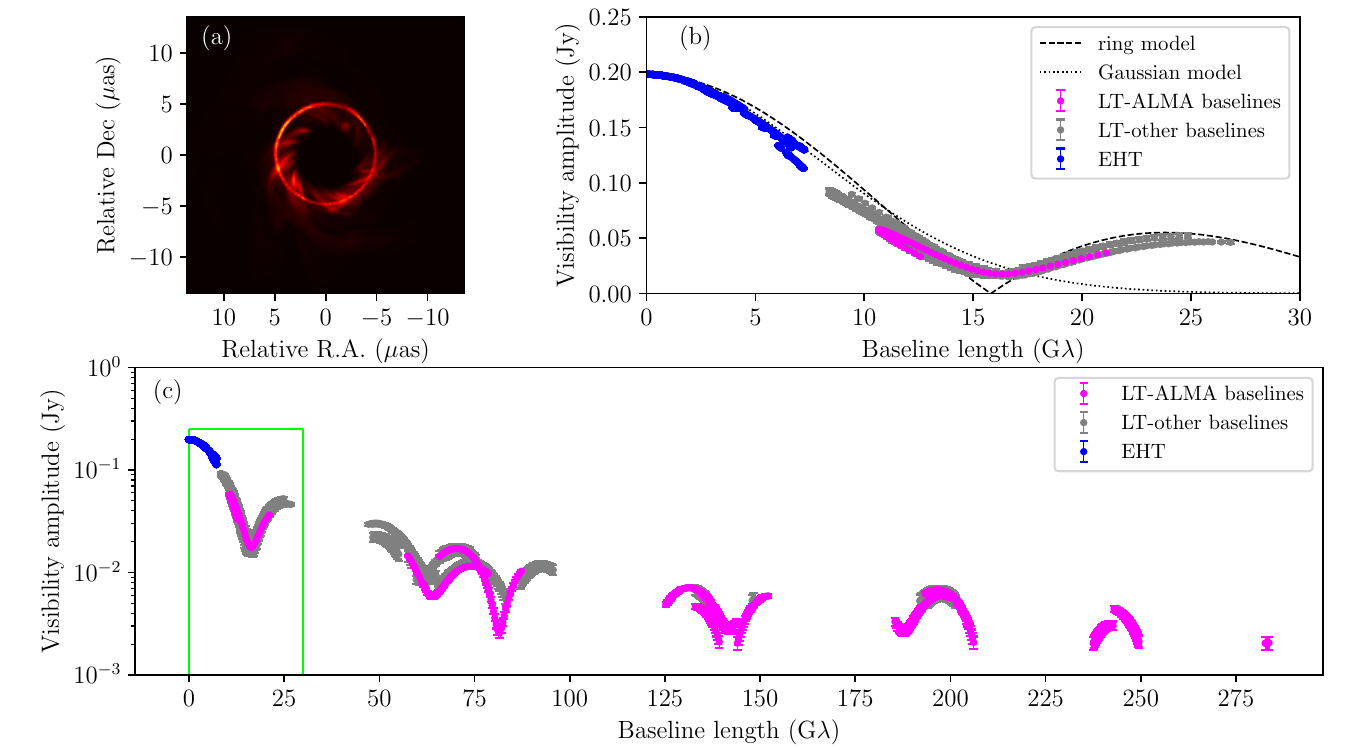}
    \caption{Simulated visibility amplitudes for M104, based on a GRMHD model, observed by the EHT and a 5 m lunar-based telescope. 
    \textbf{(a)} Black hole image from GRMHD simulation \citep{Zhang2024AA687.88} rescaled to M104's predicted shadow size (9.7 $\mu$as) and 230\,GHz flux density (0.198 Jy). 
    \textbf{(b)} Simulated visibility amplitude with thermal noise (0-30 G$\lambda$). Blue, magenta, and gray points denote EHT baselines, lunar–ALMA baselines, and lunar–EHT other-station baselines, respectively. Dashed and dotted curves show ring model ($\theta=9.7\ \mu$as, $w=3.9\ \mu$as) and Gaussian model (FWHM=9.7 $\mu$as). \textbf{(c)} Same as panel \textbf{(b)}, extended to 0-300 G$\lambda$; lime box marks the range shown in \textbf{(b)}.}
    \label{fig:synthetic_data}
\end{figure*}

Figure~\ref{fig:synthetic_data} presents the results. Panel (a) shows the input GRMHD image after rescaling. Panels (b) and (c) display the simulated visibility amplitudes over baseline lengths of 0–30 G$\lambda$ and 0-300 G$\lambda$, respectively.  The simulated data themselves show a clear null at $\sim$16 G$\lambda$, consistent with the $15.8$ G$\lambda$ predicted by the geometric ring model. Comparing the data with the ring model ($\theta=9.7\ \mu$as, $w=3.9\ \mu$as) and the Gaussian model (FWHM=9.7 $\mu$as), the ring model fits well whereas the Gaussian model does not (panel b). This demonstrates that the inclusion of Moon–Earth baselines enables a clear detection of the first null and a robust discrimination between ring and Gaussian structures.

\subsection{Photon ring detectability}\label{sec:photon_ring}

In addition, we assume that the gaps in the Moon–Earth ($u,v$) coverage can be filled by deploying multiple space-based telescopes, enabling each candidate to be observed with continuous baselines spanning from Earth to the Moon. Then we analyze the detectability of photon rings for the 31 SMBH candidates. 

Photon rings form a sequence labeled by $n=1,2,3,\dots$, where $n$ denotes the number of 1/2 times photons orbit the black hole before escaping, due to strong gravitational lensing \citep[e.g.,][]{Johnson2020SciA6.1310}. Higher-order photon rings are progressively thinner and fainter. In terms of visibility amplitudes, short baselines primarily probe the $n=0$ component (the direct image), whose emission is dominated by the accretion flow and jet. As the baseline increases, a thin ring dominated by $n=1$ emission becomes resolvable, and at even longer baselines the $n=2$ component begins to dominate the detected visibility signal.

\begin{figure*}
    \centering
    \includegraphics[width=0.98\linewidth]{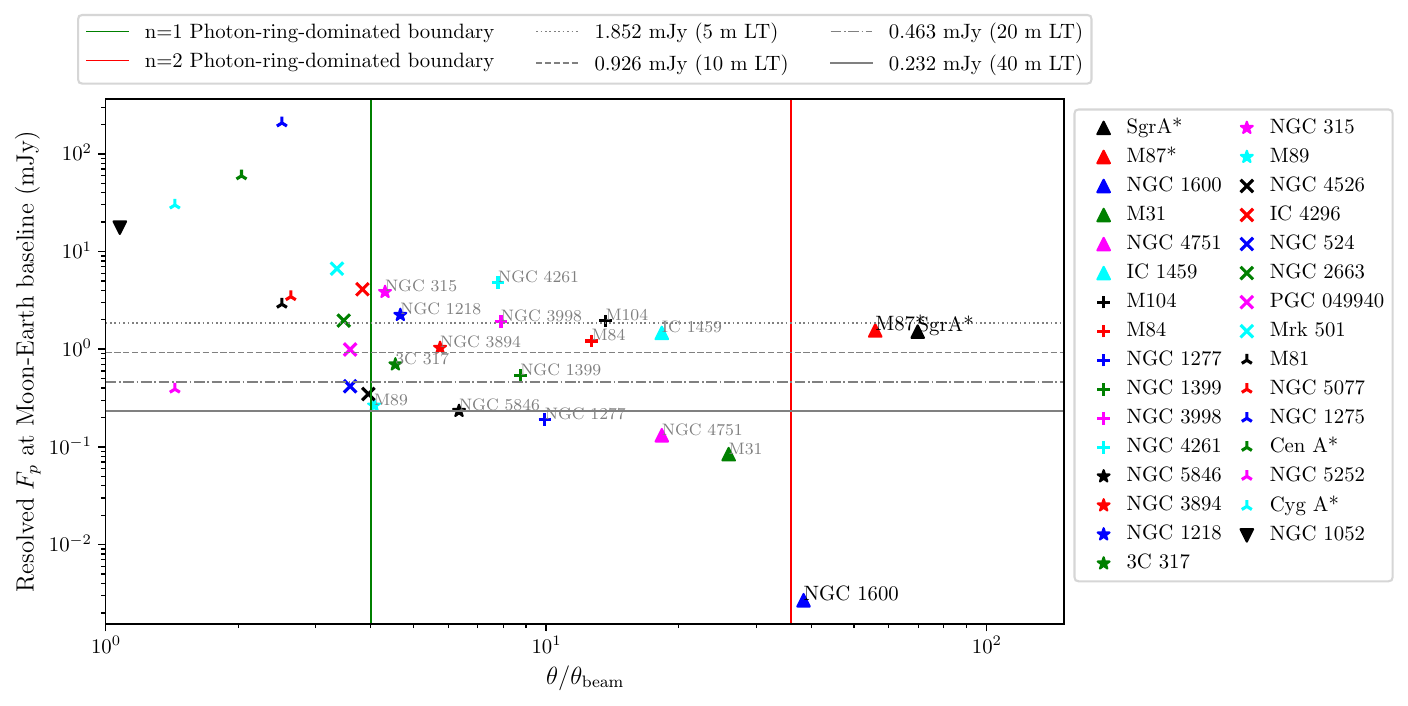}
    \caption{Photon ring detectability for 31 SMBH candidates. The x-axis shows the resolvability metric $\theta/\theta_{\rm beam}$, with vertical lines at 4 (green; $n=1$ photon-ring-dominated boundary) and 36 (red; $n=2$ photon-ring-dominated boundary). The y-axis shows the resolved photon-ring flux density $F_p$ (visibility amplitude peak) at the maximum Moon–Earth baseline, derived from a ring model with total flux density $0.1\,F_{\rm 230,\mathrm{GHz}}$, diameter $\theta$ and width $0.025\,\theta$. The horizontal lines indicate the baseline sensitivity of ALMA and the lunar-based telescope with a diameter of 5 m, 10 m, 20 m and 40 m (see Tab.~\ref{tab:sensitivity}). }
    \label{fig:photon_ring_discuss}
\end{figure*}

\cite{Johnson2020SciA6.1310,Johnson2024SPIE13092E..2DJ} define the transition between disk/jet-dominated and photon-ring-dominated regimes using the decay of visibility amplitude with baseline length, based on GRMHD simulations of M87*. For M87*, this transition occurs at $\sim$20 G$\lambda$, corresponding to $\sim$25\% of the ring diameter 42$\mu$as measured by EHT. The transition between $n=1$- and $n=2$-dominated regimes for M87* occurs at $\sim$180 G$\lambda$, 9 times larger than previous transition value (see Fig. 4 in \citealt{Johnson2020SciA6.1310}). We generalize these criteria to other candidates by adopting $\theta/\theta_{\rm beam} \geq 4$ as the condition for resolving $n=1$ photon-ring-dominated emission (right of the green line in Fig.~\ref{fig:photon_ring_discuss}), and $\theta/\theta_{\rm beam} \geq 36$ as the condition for resolving $n=2$ photon-ring-dominated emission (right of the red line). Here, $\theta$ and $\theta_{\rm beam}$ are defined in Eqs.~\eqref{eq:ring_size} and \eqref{eq:theta_beam}, respectively.

We further assume that the photon ring contributes 20\% of the compact 230\,GHz flux density, following \cite{Johnson2024SPIE13092E..2DJ}. For M87*, the compact flux density at 230\,GHz is $\sim$0.5 Jy, compared to a total flux density of $\sim$1 Jy (single-dish flux density, including unresolved jet emission \citealt{EHTC2024AA681.79}). Based on this, we assume that the compact flux density contributes 50\% of the total flux density. Thus, the photon ring contributes $\sim$10\% of the total 230\,GHz flux density $F_{\rm 230\,GHz}$. 
We model the photon ring with a total flux density of $0.1\,F_{\rm 230\,\mathrm{GHz}}$, a diameter $\theta$, and a width of $0.025\,\theta$ \citep[5\% of the radius;][]{Broderick2022ApJ935.61}. From this model, we compute the corresponding visibility amplitudes and identify the peak amplitude at baselines closest to the maximum Moon–Earth baseline. This value is then compared with the Moon–Earth VLBI baseline sensitivity threshold (horizontal lines in Fig.~\ref{fig:photon_ring_discuss}) to assess detectability.

Figure~\ref{fig:photon_ring_discuss} summarizes the results. In terms of angular resolution, Sgr~A*, M87*, NGC~1600 can resolve the n=2 photon-ring-dominated region, while 18 candidates can resolve the $n=1$ photon-ring-dominated emission region, including the three $n=2$ ring resolvable sources.
Including sensitivity constraints, 14 are sufficiently bright to be detected the $n=1$ photon ring region: 
M104, NGC~3998, NGC~4261, NGC~1218, NGC~315 require a 5 m telescope; 
Sgr A*, M87*, IC~1459, M84, NGC~3894 require 10 m; 
NGC~1399, 3C~317 require 20 m; 
and NGC~5846, M89 require 40 m.
The $n=2$ photon-ring signal, which is approximately an order of magnitude fainter than the $n=1$ ring, makes detection challenging even with a 40 m lunar-based telescope.

Our results are consistent with \cite{Johnson2020SciA6.1310}, showing that Sgr~A* and M87* can resolve the $n=2$ photon ring with Moon-Earth baselines. 
Among the other sources with a detectable $n=1$ photon ring region, seven (M104, NGC~3998, NGC~4261, NGC~315, IC~1459, NGC~3894, and 3C~317) are also BHEX targets \citep{Johnson2024SPIE13092E..2DJ}. 
This suggests a stepwise progression in VLBI capability, from ground-based arrays to Earth-orbital baselines (e.g., BHEX), and then to Moon–Earth baselines. For Sgr~A* and M87*, this means advancing from horizon-scale imaging to the $n=1$ photon ring and further to the $n=2$ photon ring. For other sources (e.g., M104, IC~1459), it enables imaging from jet-dominated scale to horizon-scale and ultimately to the $n=1$ photon-ring scale.

\section{conclusion and discussion}\label{sec:conclusion}

We evaluate 31 black hole shadow candidates using simulated Moon–Earth VLBI observations combining a lunar-based telescope with the EHT. After assessing five potential lunar sites, we find that an equatorial location enables all candidates to be observed above a 15$^\circ$ elevation limit. Assuming a telescope located at the lunar antipode (180$^\circ$ E, 0$^\circ$ N), we analyze shadow detectability for a lunar-based telescope of 5 m, 10 m, 20 m, and 40 m, adopting 50\% aperture efficiency, a system temperature of 100 K, 2-bit sampling, an 32\,GHz bandwidth, and a 10 min integration time. Modeling the sources as uniform rings and applying both resolvability and sensitivity criteria in the visibility domain, we identify six high-priority targets whose shadows are detectable with a 40 m lunar-based telescope: M104, NGC 524, PGC 049940, NGC 5077, NGC 5252, and NGC 1052.

The six shadow-detectable candidates are selected primarily due to their favorable sky positions. Their near coplanarity with the Moon’s orbit produces an almost linear Moon–Earth ($u,v$) coverage, enabling projected baselines from near-Earth to lunar distance and thus providing dense samplings of visibility along a certain direction. M104 is the most promising target: its large predicted shadow size ($\sim$10~$\mu$as) is resolvable at relatively short projected baselines and its high flux density ($\sim$200~mJy) enables detection even with a 5~m lunar-based telescope. NGC 524, PGC 049940, and NGC 5077 have smaller black hole shadow sizes ($\sim$2~$\mu$as) and require 20 m, 10 m, 5 m, antennas respectively. NGC~5252 and NGC~1052 have even smaller shadows ($\sim$1~$\mu$as) and are more sensitive to the uncertainties in black hole mass and distance (see Appendix~\ref{sec:uncertainties}); while NGC~1052 is less challenging due to its high flux density ($\sim$350~mJy, 5 m antenna is sufficient), NGC~5252 is significantly fainter and thus more challenging to detect, even with a 40 m antenna. 

Our results rely on an idealized ring model. Owing to the Moon–Earth geometry, visibility detections are only well resolved along a single direction in the ($u,v$) plane, aligned with the projected lunar orbit. The lack of symmetry in coverage can introduce degeneracies between a ring and alternative source morphologies, such as a pair of compact Gaussian components. Furthermore, the black hole shadow image structure may become less ring-like when the inclination is very high. In such a case, the accretion disk is viewed edge-on, and the Doppler effect makes one side of the ring significantly brighter than the other. In addition, horizon-scale variability may temporarily distort the source morphology away from a ring-dominated structure. In such cases, null detection becomes more sensitive to baseline orientation, posing an even greater challenge for the highly asymmetric baseline coverage. Placing space telescopes on highly inclined orbits relative to the lunar plane could help mitigate this limitation. More generally, deploying multiple space telescopes on suitably designed orbits would provide denser and more symmetric ($u,v$) coverage, enabling sub–microarcsecond imaging of black hole shadows for a broader range of candidates. Optimizing such a configuration is left for future work. 

We also assess the photon ring detectability of the 31 SMBH candidates, leaving aside the baseline coverage issue (e.g., assuming gaps can be filled by space telescopes between Moon and Earth). Sgr~A*, M87* and NGC 1600 can resolve the $n=2$ photon ring dominated region at Moon–Earth baselines, consistent with \cite{Johnson2020SciA6.1310}, but the sensitivity is insufficient even with a 40 m lunar-based telescope. The $n=1$ photon ring dominated region is detectable for 14 candidates with antenna size up to 40 m, five of which (M104, NGC~3998, NGC~4261, NGC~1218, and NGC~315) require only a 5 m lunar-based telescope. These results demonstrate that Moon–Earth VLBI offers a pathway to strong-field gravity studies at sub-microarcsecond angular resolution across diverse AGN sources.

This work represents a first step toward assessing black hole shadow detectability with Moon–Earth VLBI. More practical consideration, including capability parameter optimization (e.g., bandwidth, integration time), detailed observing strategies, and technical implementation (e.g., data downlink, whether the aperture is realized by synthesizing small dishes or a single large dish), remain to be explored in future work.

\section*{Acknowledgements}
We thank the referee for the detailed and constructive comments, which have significantly improved this work. We thank Dr. Dhanya G. Nair for comments. This work is supported by the National Science and Technology Major Project (2024ZD1100601),  the National Natural Science Foundation of China (Grant No. 12325302, 11933007), the Key Research Program of Frontier Sciences, CAS (grant no. ZDBS-LY-SLH011), and the Shanghai Pilot Program for Basic Research, Chinese Academy of Sciences, Shanghai Branch (JCYJ-SHFY-2021-013). LL is supported by the International Collaboration Project on Space VLBI, \textsl{Key Technologies in Space VLBI Data Processing} ( Project No. ZA06). YM is supported by the National Key R\&D Program of China (grant no. 2023YFE0101200), the National Natural Science Foundation of China (grant no. 1273022, 12511540053), and the Shanghai municipality orientation program of basic research for international scientists (grant no. 22JC1410600).

\section*{Data Availability}
The data generated in this work will be shared on reasonable request to the corresponding author.




\bibliographystyle{mnras}
\bibliography{example} 




\appendix

\section{Minimal projected Moon-Earth baseline distribution}
\label{sec:minimal_baseline}

\begin{figure*}
    \centering
    \includegraphics[width=0.98\linewidth]{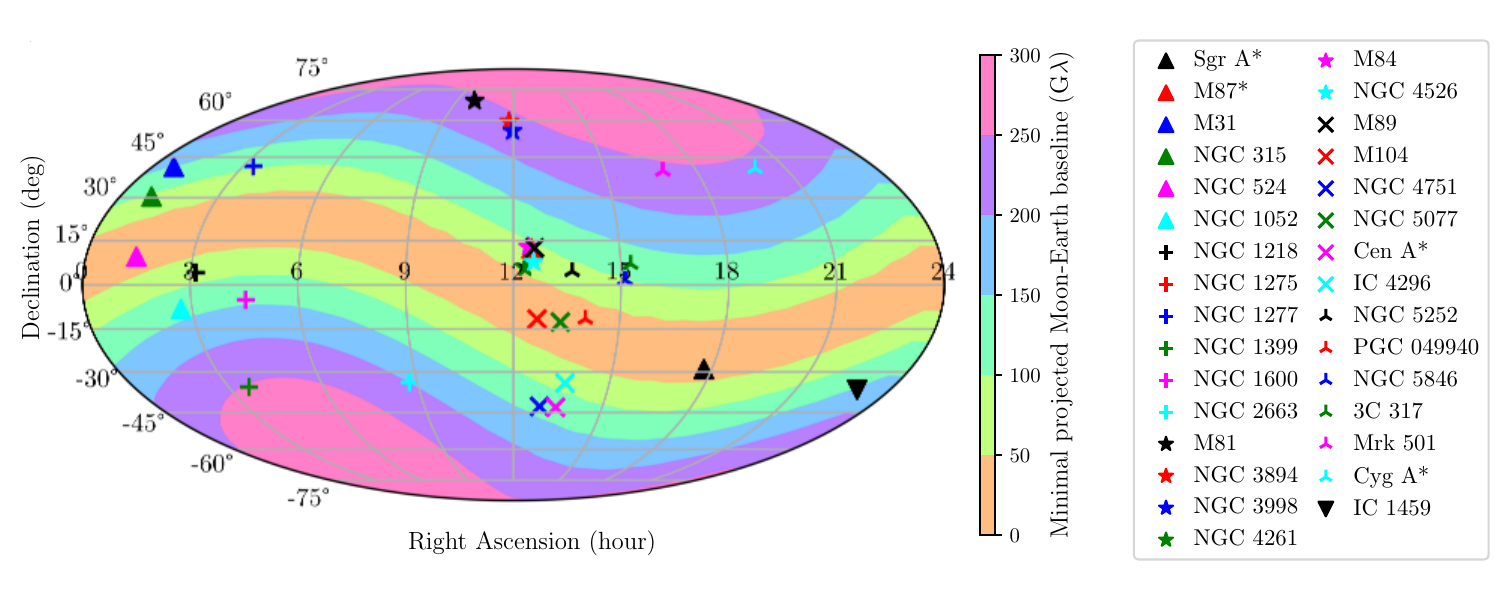}
    \caption{Distribution of the minimal projected Moon-Earth baseline on the celestial sphere, with marked positions of SMBH candidates. The sinusoidal pattern reflects the geometric relationship between the source position and the Moon's orbital path (with a 5.15$^\circ$ mean inclination relative to the equator); sources located near this orbital path exhibit smaller minimal projected baselines, corresponding to the linear ($u,v$) coverage cases presented in Fig.~\ref{fig:uv_coverage}.  }
    \label{fig:BLmin}
\end{figure*}

We extract the minimal projected Moon–Earth baselines from the ($u,v$) coverage simulation (Section~\ref{sec:VLBIsimulation}) and plot their distribution on the celestial sphere (Figure~\ref{fig:BLmin}). The sinusoidal pattern reflects the geometry of the Moon's orbit, which is inclined at a mean value of 5.15$^\circ$ relative to the Earth's equator. Sources near the Moon's orbital path generate smaller minimal projected baselines, corresponding to the linear ($u,v$) coverage cases in Figure~\ref{fig:uv_coverage}. In contrast, sources with higher declinations produce larger minimal projected baselines, which correspond to the circular ($u,v$) coverage cases.

\section{detectability uncertainties from prior of mass and distance}\label{sec:uncertainties}

\begin{figure*}
    \centering
    \includegraphics[width=0.98\linewidth]{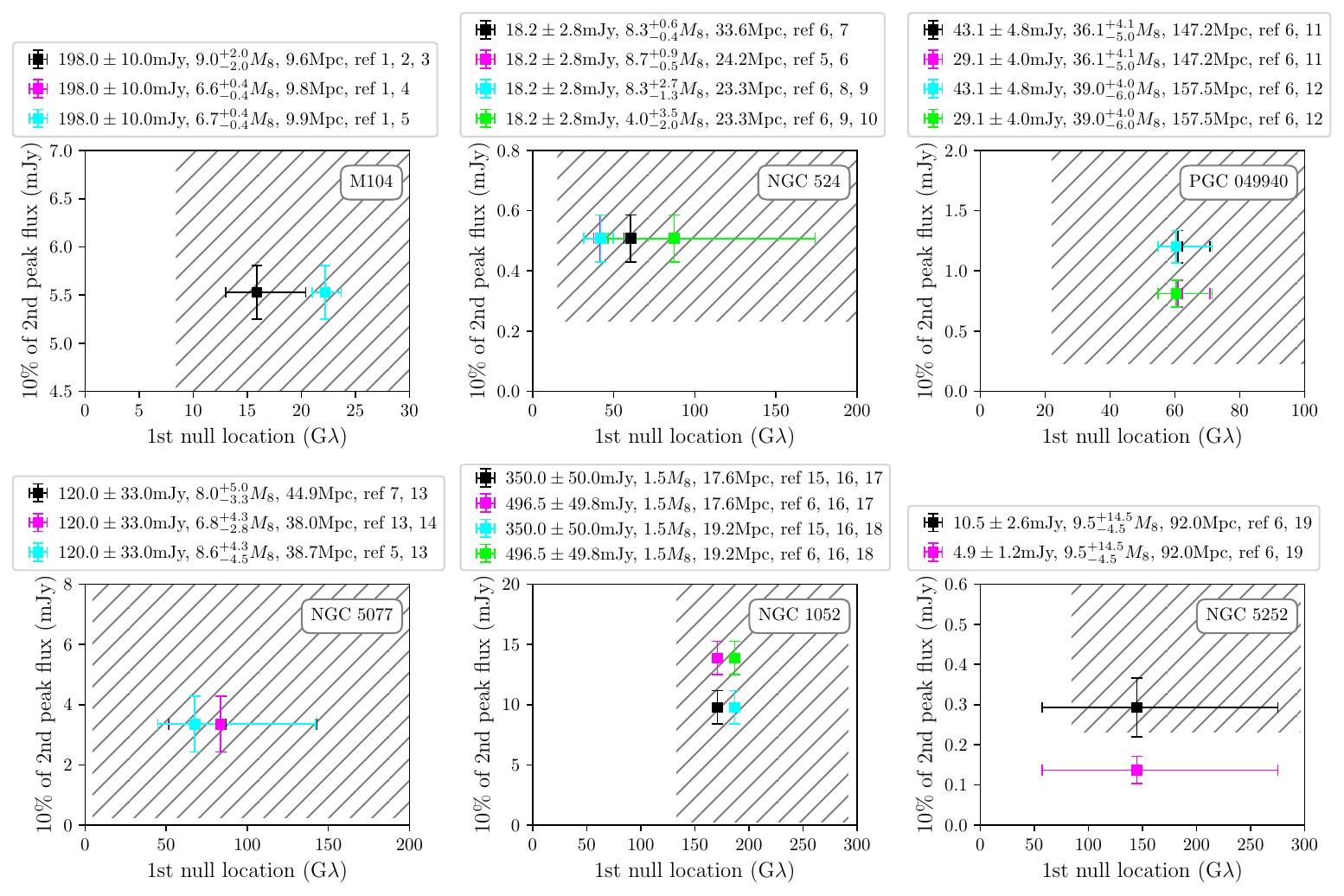}
    \caption{Detectability uncertainties for the six shadow-detectable candidates. The x- and y-axes show the first null location and 10\% secondary peak flux density of the ring model visibility amplitudes. Colored markers with error bars indicate values and uncertainties derived from different assumptions for the 230\,GHz flux density, black hole mass, and distance (see legend; black markers denote the values adopted in this work; $M_8 = 10^8\,\rm{M}_\odot$). The hatched region marks the detectable parameter space, constrained by the projected Moon–Earth baseline range (x-axis) and the 0.232 mJy sensitivity limit (y-axis) for a 100 m lunar-based telescope.\\
    References: 1 ACSC band 6 data; 
    2 \protect\cite{Menezes2015ApJ808.27}; 
    3 \protect\cite{McQuinn2016AJ152.144}; 
    4 \protect\cite{Jardel2011ApJ739.21}; 
    5 \protect\cite{Gultekin2019ApJ871.80}; 
    6 \protect\cite{Lo2023ApJ950.10}, with flux densities of PGC 049940 and NGC 5252 observed from multiple epochs; 
    7 \protect\cite{Sani2011MNRAS413.1479}; 
    8 \protect\cite{Krajnovic2009MNRAS399.1839}; 
    9 \protect\cite{Tonry2001ApJ546.681}; 
    10 \protect\cite{Smith2019MNRAS485.4359}, 
    11 \protect\cite{DallaBonta2009ApJ690.537}; 
    12 \protect\cite{McConnell2011Natur480.215}; 
    13 \protect\cite{Chen2023ApJ951.93}; 
    14 \protect\cite{deFrancesco2008AA479.355}; 
    15 \protect\cite{Baczko2024AA692.205}; 
    16 \protect\cite{Woo2002ApJ579.530}; 
    17 \protect\cite{Kameno2020ApJ895.73}; 
    18 \protect\cite{Tully2013AJ146.86}; 
    19 \protect\cite{Capetti2005AA431.465}.  
    }
    \label{fig:uncertainties}
\end{figure*}

The detectability analysis in section~\ref{sec:criteria} depends on physical parameters of the candidate: the 230\,GHz flux density ($F_{\rm 230\,GHz}$), black hole mass ($M_{\rm BH}$), and distance ($D_{\rm BH}$). The location of the first null in the ring model visibility scales with $D_{\rm BH}/M_{\rm BH}$, such that larger masses or smaller distances shift the null to shorter baselines. Secondary peak flux density scales with $F_{\rm 230\,GHz}$.

Using different physical parameters and uncertainties for the six shadow-detectable candidates (Sect.~\ref{sec:best}), we compute the corresponding first null locations and 10\% secondary peak flux densities and show them in Fig.~\ref{fig:uncertainties}. M104, NGC~524, PGC~049940, and NGC~5077 are robustly detectable, with both indicators and their uncertainties lie entirely within the observable range. 

In contrast, the detectability of NGC~1052 and NGC~5252 is less certain. For NGC~1052, the sensitivity criterion is satisfied, but the absence of reported mass uncertainties precludes assessment of whether shifts in the first null location could move it outside the resolvable baseline range. For NGC~5252, large mass uncertainties imply that the first null location could fall outside the baseline coverage, and its potentially low 230\,GHz flux density \citep[$4.9 \pm 1.2$ mJy, magenta maker;][]{Lo2023ApJ950.10} causes it to drop below the sensitivity threshold, meaning it requires a lunar-based telescope larger than 40 m.

\section{Uncertainties from brightness distribution asymmetry}
\label{sec:asymmetry}

\begin{figure*}
    \centering
    \includegraphics[width=0.9\linewidth]{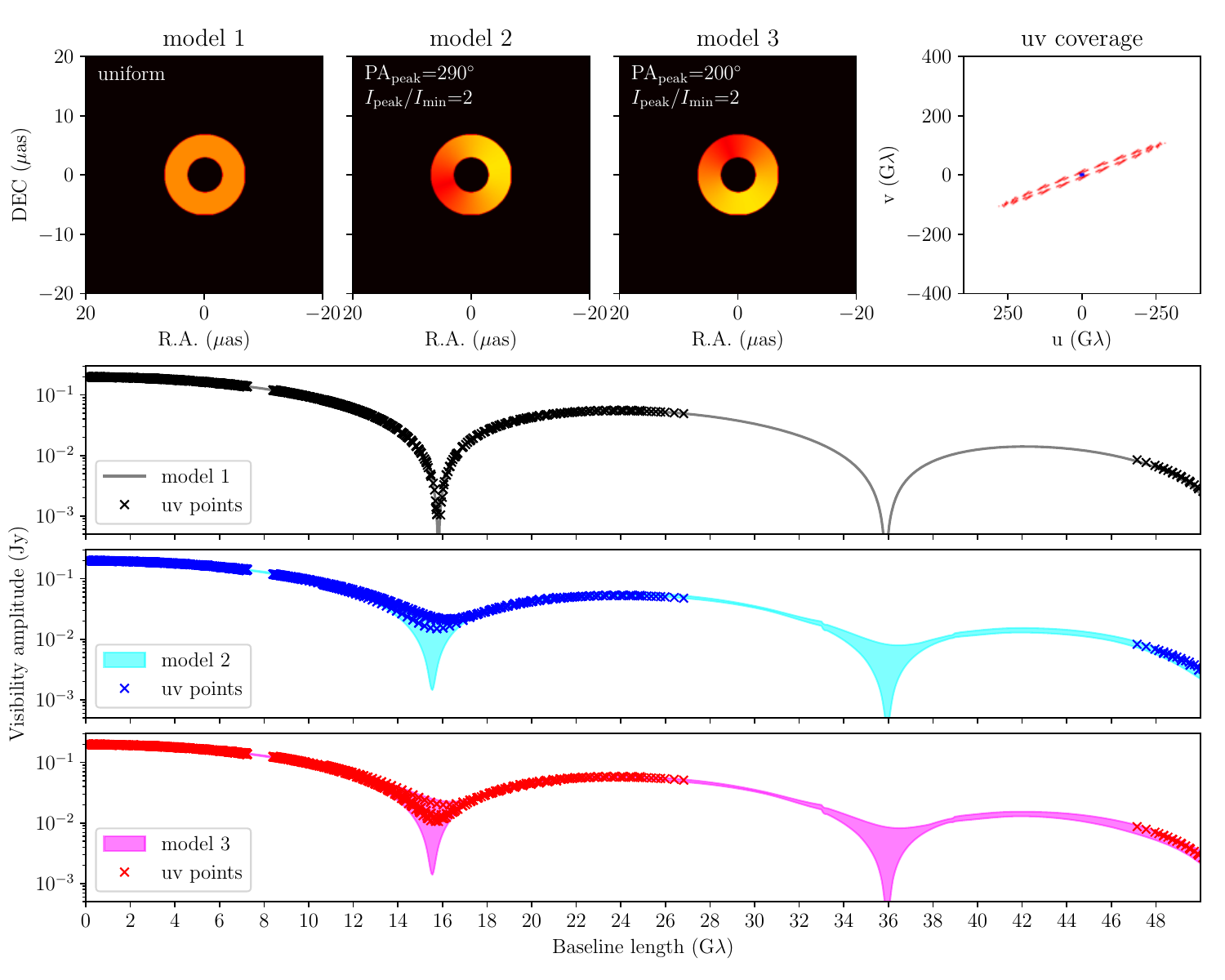}
    \caption{Comparison of first-null detectability for M104 under different azimuthal brightness assumptions. Model 1 assumes a uniform ring; models 2 and 3 include azimuthal asymmetry with a peak-to-minimum brightness ratio of 2, with peaks at 290$^\circ$ (aligned with the Moon–Earth baseline) and 200$^\circ$ (perpendicular), respectively. Top row: model images and (u, v) coverage. Rows 2 to 4: visibility amplitudes of baseline length from 0 to 50 G$\lambda$, with model 1 shown as a gray solid curve and models 2–3 as cyan and magenta shades. Crosses mark sampled (u, v) points for each model (black, blue, red). }
    \label{fig: asymmetry}
\end{figure*}

In addition to the idealized uniform ring adopted in this work, we briefly examine the impact of azimuthal brightness asymmetry on the detectability of the first null with Moon–Earth VLBI. Using M104 as a representative example, we extend the uniform ring model (model 1 in Fig.~\ref{fig: asymmetry}) by introducing a Gaussian modulation in position angle with a peak-to-minimum brightness ratio of 2. Two representative configurations are considered: a brightness peak aligned with the Moon–Earth baseline (290$^\circ$) and one perpendicular to it (200$^\circ$), corresponding to models 2 and 3. We compute the sampled visibility amplitudes from the Fourier-transformed visibilities of model images 1–3 using the Moon-Earth VLBI $(u, v)$ coverage, and compare their differences around the first null.

We find that, while brightness asymmetry modifies the detailed visibility structure around the first null, the null location itself remains unchanged. The primary effect of brightness asymmetry is to broaden the visibility amplitude distribution near the null, so that different baseline orientations yield a range of values rather than a sharp zero crossing. In this sense, the null becomes "shallower" and less distinct. Specifically, for sampled baseline lengths in the first null region, the visibility amplitudes increase from $\lesssim$0.1\%
of the total flux density in the symmetric case to 
$\sim$8\% and $\sim$5\% in the two asymmetric cases.

Importantly, this effect does not alter our main conclusion: using the baseline length at which the first null occurs to constrain ring size, remains robust under reasonable levels of asymmetry. The detailed visibility amplitude behavior near the null, however, depending on specific brightness distribution, which in turn is governed by the source physics (e.g., inclination, accretion flow structure, relativistic radiation). A comprehensive treatment of such effects would therefore require more realistic, source-dependent modeling and is beyond the scope of this work. We defer a detailed investigation of asymmetry-induced effects to future studies.

\bsp	
\label{lastpage}
\end{document}